\documentclass[floats,floatfix,showpacs,amssymb,prd,twocolumn,superscriptaddress,nofootinbib,longbibliography,reprint]{revtex4-1}
\usepackage{amssymb,amsmath,verbatim,mathtools,needspace,enumitem,etoolbox,graphicx,physics,microtype,afterpage,xspace,tabularx,lmodern,multirow}
\usepackage{blindtext}
\usepackage{graphicx}
\usepackage{dcolumn}
\usepackage{xcolor}
\usepackage{bm}
\usepackage{multirow}
 
\usepackage[dvipsnames, usenames]{xcolor}
\definecolor{linkcolor}{rgb}{0.65, 0.16, 0.16}
\usepackage[unicode, colorlinks=true, linkcolor=linkcolor, citecolor=linkcolor, filecolor=linkcolor, urlcolor=linkcolor, linktocpage, breaklinks]{hyperref}
\usepackage{calligra}
\usepackage[T1]{fontenc}
\usepackage{egothic}
\usepackage[T1]{fontenc}
\newfont{\rsfsten}{rsfs10 scaled 1200}
\newfont{\rsfsseven}{rsfs10 scaled 1200}
\newfont{\rsfsfive}{rsfs10 scaled 1200}
\usepackage{epsfig}
\usepackage{units}
\usepackage[utf8]{inputenc}
\usepackage{hyperref}
\usepackage{tabularx}
\newcommand{\be}{\begin{equation}}
\newcommand{\ee}{\end{equation}}
\newcommand{\bea}{\begin{eqnarray}}
\newcommand{\eea}{\end{eqnarray}}

\usepackage{comment}

\usepackage{orcidlink}

\begin{document}
\title{Binary Black Holes population synthesis based on the current LVK observations}

\author{Mehdi El Bouhaddouti\,\orcidlink{0009-0001-1299-0879}}
\email{melbouhaddouti@oakland.edu}

\author{Ilias Cholis\,\orcidlink{0000-0002-3805-6478}}
\email{cholis@oakland.edu}

\author{Muhsin Aljaf\,\orcidlink{0000-0002-1251-4928}}
\email{muhsinaljaf@oakland.edu}

\affiliation{Department of Physics, Oakland University, Rochester, Michigan, 48309, U.S.A}

\begin{abstract}

The ongoing observations from ground based gravitational-wave observatories have led to the detection
of more than a hundred merger events between black holes. 
We use the LIGO-Virgo-KAGRA (LVK) observations from 2015 to early 2024, to test the population synthesis of these
merging binaries; which will allow us to probe the formation mechanisms and environments of these black holes.  
We test if the current sample of binary black holes can be explained only by the merger of black holes 
coming from the collapse of the cores of massive stars, i.e. as just first generation black holes merging with 
each other. Those black holes' masses will roughly follow a power-law distribution. We also test if in addition to the merger between first generation black holes, there is evidence for 
a second population of black hole binaries in which at least one the binaries' members is the product of an earlier 
merger between black holes. 
These binaries are typically referred to as signals of hierarchical mergers. 
Such a population can possibly explain the observation of very massive black hole binaries by the LVK collaboration. 
We find that the LVK observations give a statistical preference in log-likelihood of up to $- 2 \Delta ln\mathcal{L} = -150$ or in log-Bayes factor of up to $ln\textrm{BF} = 71$, for the full sample of black hole 
binaries originating from a combination of black holes following a power-law distribution and black holes from hierarchical mergers. The ratio of black holes following a power-law mass-distribution to a mass-distribution expected from hierarchical mergers is found to be as high as one-to-one. 
We also consider that some of the LVK black hole merging binaries are the result of primordial black holes (PBHs), merging 
inside dark matter halos and in the intergalactic medium. Adding a third population is preferred. 
Compared to a simple first generation population we get $- 2 \Delta ln\mathcal{L} = -216$ and $ln\textrm{BF} = 97$. 
We find that a few $\%$ of the observed LVK events can be PBH merging binaries. Depending on the exact assumptions on the formation of PBH binaries, this translates to stellar-mass range PBHs accounting for a fraction $f_{\textrm{PBH}} \simeq 3\times 10^{-3}-2\times 10^{-2}$ of the observed 
dark matter abundance. 
Such a population
of PBHs in combination with hierarchical mergers can best explain the observed LVK events with black hole masses in the range 
of $\simeq 30-45 \, M_{\odot}$. 

\end{abstract}

\maketitle

\section{Introduction} \label{sec:intro}
Since the first detection in 2015 of the merger event of two compact objects \cite{PhysRevLett.116.061102}, the Laser Interferometer Gravitational-wave Observatory (LIGO) \cite{2015CQGra..32g4001L}, in combination with Virgo \cite{T_Accadia_2012} and the Kamioka Gravitational Wave Detector (KAGRA) \cite{KAGRA:2020tym}, have completed three observation runs (O1–O3) \cite{PhysRevX.9.031040, PhysRevX.13.011048}, while results from the ongoing fourth run (O4) have recently been released in Ref.~\cite{LIGOScientific:2025slb}. The number of binary black holes (BBHs) in the catalog of binary compact object mergers detected by the LIGO-Virgo-KAGRA collaboration (LVK), with a false alarm rate (FAR) of $< 1 \; \textrm{yr}^{-1}$, has reached 153 \cite{LIGOScientific:2025slb}. Each event in the Gravitational Wave Transient Catalog (GWTC) has an estimate of the primary mass $m_1$, secondary mass $m_2$, and redshift $z$, among other quantities. This allows the study of the mass and redshift distributions of the black holes (BHs) that form binaries, giving a merger signal in the gravitational-wave detectors.
Moreover, one can use the reported $m_{1}$, $m_{2}$ and $z$ information and the LVK reported noise curves to evaluate signal to noise ratios (SNR). Using the reported LVK merger events there are 159 BBHs with an SNR $>8$ and $m_{2} > 4 \, M_{\odot}$. We use those events in this analysis. 

Previous studies of the LVK data have also discussed the origin of the BBHs, including the possibility that the observations 
probe multiple populations of BBHs (see e.g. \cite{Hutsi:2020sol, Ng:2020qpk, Tiwari:2021yvr, DeLuca:2021wjr, Franciolini:2021tla, Karathanasis:2022rtr, Liu:2022iuf, Park:2022mez, Ulrich:2024nez, Bouhaddouti:2024ena, Li:2024rmi, Gennari:2025nho, LIGOScientific:2025pvj}). 
We consider three populations for the origin of BBHs: i) binaries following a power-law mass (PL) distribution, as for instance made of first generation BHs (of stellar origin) 
 \cite{Fryer:1999ht, 2010ApJ...715L.138B, 2010ApJ...714.1217B, 2012ApJ...749...91F, Rodriguez:2016kxx, Mirabel_2017}, ii) binaries made of BHs resulting from previous mergers, known as hierarchical mergers 
 (HM)~\cite{Antonini:2016gqe, Fishbach:2017dwv, Gerosa:2017kvu, Kovetz:2018vly, Antonini:2018auk, Kritos:2020wcl, Purohit:2024zkl, Fragione:2023kqv}, and 
 iii) binaries made of primordial black holes (PBHs)~\cite{Sasaki:2016jop, Bird:2016dcv, Kovetz:2016kpi, Kovetz:2017rvv,  Kavanagh:2018ggo, Clesse:2020ghq, Franciolini:2022ewd, DeLuca:2022bjs, Carr:2023tpt, Raidal:2024bmm, Aljaf:2024fru, Carr:2025kdk}. 
 
First generation BHs can be expected to have an approximately power-law mass distribution, starting from a minimum mass of $\simeq 4 \, M_{\odot}$ with or without a high-mass cut-off, depending on the contribution of population III stars~\cite{Tanikawa:2024mpj, Wiktorowicz:2019dil}. 
Such a population of BHs may have a distribution similar to that of their progenitor stars. 
However, the exact metallicity distribution of the progenitor stars and the environment where these BHs are hosted, may also cause some deviations from a simple power-law on the mass-distribution of the observable 
by gravitational waves remnant compact objects (see e.g.~\cite{2012ApJ...749...91F, Mapelli:2012vf, Morscher:2014doa, Ziosi:2014sra, Kimpson:2016dgk}).  
For stars with a mass larger than $1 \, M_\odot$, including massive enough to form BHs after their collapse, 
their initial mass follows a power-law distribution
of $dN/dM \mathtt{\sim} M^{-2.3\pm0.7}$ \cite{Kroupa:2000iv}.
For simplicity, we will consider such a power-law distribution to model first generation BHs, allowing for the power-law index to be set by the LVK data. 
Due to the pair-instability mass gap~\cite{Woosley:2021xba}, the number of first generation BHs is expected to drop 
significantly after some value of mass. For the first generation BHs, staring at $4 \ M_{\odot}$, we test both a simple power-law mass spectrum, and a mass spectrum described by a power-law with an exponential cutoff at $40M_\odot$. 

By HM binaries, we refer to the merger of BHs where the most massive member  $m_{1}$, is at least a second generation BH. 
The presence of HM events in the gravitational-wave observations has been discussed as early as 2017 
\cite{Fishbach:2017dwv, Gerosa:2017kvu}, when only two events from the O1 run were publicly available and the O2 run 
was still ongoing. Using events from runs O1, O2 and O3, Ref.~\cite{Li:2024zwt} found that at least $10\%$ of GWTC3' s events 
are HMs, while also \cite{KAGRA:2021duu, Ulrich:2024nez, Bouhaddouti:2024ena}, found indications for a similar size 
population of events not originating from first generation BHs.
To study the HM population, we rely on the stellar cluster simulations from Ref.~\cite{Ye:2025ano}, from where we model the binaries' primary mass $m_{1}$ distribution and the ratio of secondary mass ($m_{2}$) to primary mass, $q=m_{2}/m_{1}$ distribution.

PBHs \cite{Hawking:1971ei, Zeldovich:1967lct, 1975Natur.253..251C}, may have formed in the early Universe from the collapse 
of density perturbations. In this case, they would contribute to the observed dark matter abundance, even if just as a fraction of it. 
Furthermore, mergers between such objects can give a detectable signal in gravitational-wave observations~\cite{Sasaki:2016jop, Bird:2016dcv, Carr:2016drx}. 
The most direct observable signal from PBHs is that they can contribute to the BBH merger rates and mass spectra measured 
by the LVK collaboration~\cite{LIGOScientific:2016aoc, KAGRA:2021duu, LIGOScientific:2025slb}. 
Based on the most up to date estimates on the PBH binaries merger rates of Ref.~\cite{Aljaf:2025dta}, we test if a 
PBH component can exist in the current LVK BH primary and secondary mass-spectra. This allows us to probe and set 
constrains on the PBH abundance in the stellar mass range. We derive our results in terms of the fraction of dark matter
abundance that is in PBHs $f_{\rm PBH}= \rho_{\textrm{PBH}}/\rho_{\rm dm}$, where $\rho_{\textrm{PBH}}$ is the averaged 
energy density in PBHs in the Universe and $\rho_{\rm dm}$ is the dark matter energy density of the Universe.

In Section~\ref{sec:method}, we describe the LVK data that we use to study the $m_{1}$, $m_{2}$ and redshift $z$ distributions of the 
observed BBH merges. We also describe the alternative ways in which we model the merger rates from the three populations 
of BBHs (first generation, HM, and PBH), to account for enough modeling flexibility on each of the three populations. 
In Section~\ref{sec:results}, we show how these populations of merging BBHs are constrained by 
the current LVK observations, accounting for the sensitivity of LVK at different frequencies i.e. the observation biases. 
We find that between 10 and 50 $\%$ of all the total merger rate of BBH at low redshifts is of hierarchical origin. 
Such BH binary mergers can explain the significant population of detected BBHs that contain a BH with a mass of $\geq 30 \, M_{\odot}$. 
Moreover, we find that merging PBH binaries in the stellar-mass range are allowed to contribute up to a few $\%$ of the total merger rate at $z \leq 1$. Depending on the exact assumptions on the formation of PBH binaries, this translates to a fraction of $f_{\rm PBH}= 3 \times 10^{-3} - 2\times10^{-2}$ on their abundance. 
Finally, in Section~\ref{sec:conc}, we present our conclusions and discuss how future observations
from ground-based gravitational wave observatories will further inform us on the origins of stellar-mass BHs.

\section{Methodology} \label{sec:method}

We first discuss the LVK observations that we use to study the mass and redshift distributions
of the BBHs. We then describe the modeling of the three populations of merging BBHs. 

\subsection{The LVK population of black hole binaries}
\label{subsec:pop_stud}

The two LIGO detectors at Hanford Washington and Livingston Louisiana, have a much better sensitivity compared to the 
Virgo and KAGRA detectors. Still, Virgo and KAGRA allow for an improved localization of the events. For this analysis where 
we want to use the observed $m_{1}$, $m_{2}$ and redshift $z$-distribution of the BBH mergers in order to probe the underlying 
populations of these objects, we will rely on the publicly available noise curves for the two LIGO interferometers \cite{Aasi_2015,ligo_scientific_collaboration_and_virgo_2023_8177023,ligo_scientific_collaboration_and_virgo_2025_16053484}. 

\begin{figure}[h]
\includegraphics[keepaspectratio,width=0.95\linewidth]{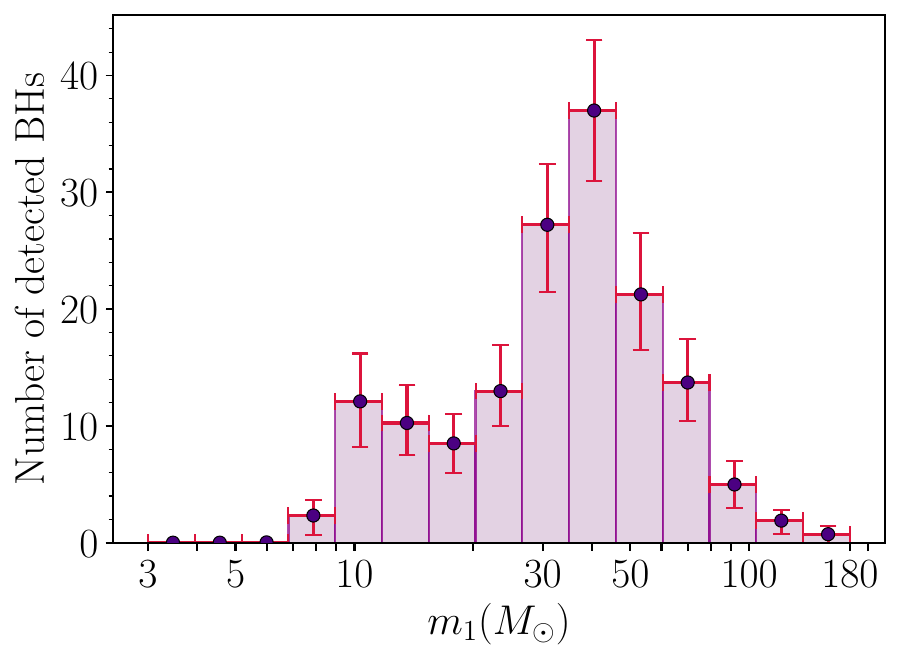} \\
\includegraphics[keepaspectratio,width=0.95\linewidth]{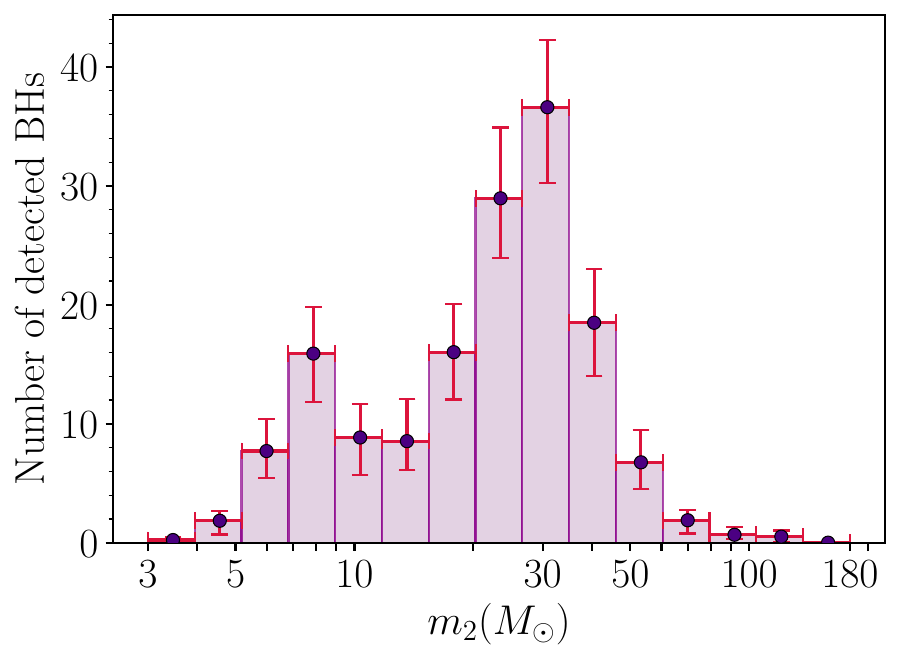} \\    
\includegraphics[keepaspectratio,width=0.95\linewidth]{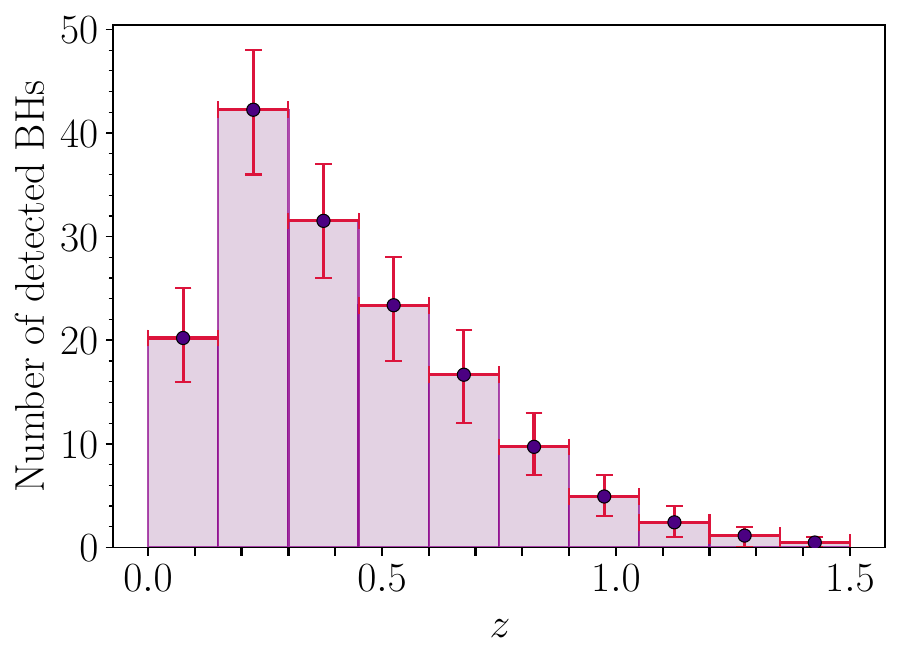}
\caption{Histogram of primary mass (top), redshift (middle), and secondary mass (bottom) distribution for BBHs, with Poisson error bars from the GWTC4 events.}
\label{fig:m_1_2_z_hists}
\end{figure}

We use the events from runs O1 to O4a, with $m_1$ and $m_2$ central values larger than $4M_\odot$ and $\text{SNR}\ge 8$. 
For each detected event, the LVK collaboration provides the posterior samples on the $m_{1}$ and $m_{2}$ masses \cite{ligo_scientific_collaboration_and_virgo_2023_8177023,ligo_scientific_collaboration_and_virgo_2025_16053484}. 
We use those distributions to make discretized histograms for $m_{1}$ and $m_{2}$ respectively. 
Of the 159 events with $m_{2}> 4 \, M_{\odot}$ and $\text{SNR}\ge 8$ there are posterior samples for the 153 BH binaries. 
We normalize these histograms  so that the sum on all bin counts is equal to 1. To obtain our representation of the LVK BBH data, we stack all the $m_1$ discretized probability density functions (PDFs) to obtain the $m_1$ distribution, and we stack the $m_2$ PDFs to obtain the $m_2$ distribution. 
These stacked distributions are shown in Fig.~\ref{fig:m_1_2_z_hists}. 
Both histograms consist of 15 bins logarithmically spaced between $3M_\odot$ and $180M_\odot$. 
Values of masses smaller than $4 \, M_{\odot}$ appear as the reported LVK posterior samples in \cite{ligo_scientific_collaboration_and_virgo_2023_8177023,ligo_scientific_collaboration_and_virgo_2025_16053484} extend to these lower masses. 
To produce the redshift distribution histogram of Fig.~\ref{fig:m_1_2_z_hists} for all LVK events, we first take for each binary its relevant redshift posterior distribution.
We then stack the redshift posterior distributions from the entire events sample and bin them in 10 linear bins up to z=1.5.

\begin{figure}
    \includegraphics[keepaspectratio,width=0.99\linewidth]{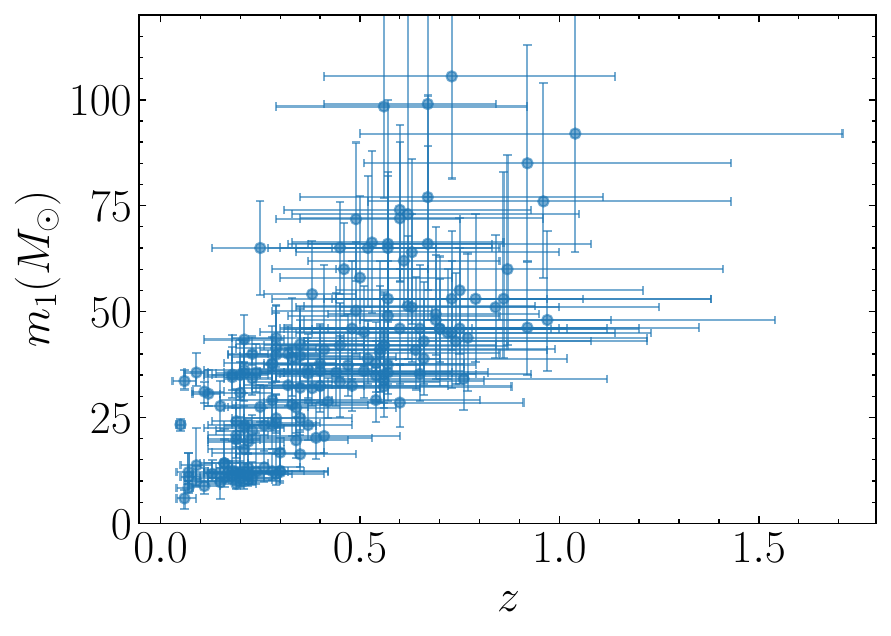}
    \caption{The primary mass of events from the O1 to O4a LVK runs, as a function of redshift with their $90\%$ credible interval ranges.}
    \label{fig:m_1_z_distr}
\end{figure}

In Fig.~\ref{fig:m_1_z_distr}, we show the $m_{1}$ vs $z$ distribution of the 153 LVK BBH events that we use in this analysis. The error-bars represent the $90\%$ credible interval reported ranges. 

\subsection{Merger rates from first generation black holes}
\label{subsec:first_generation}
First generation black holes originate from the core collapse of stars. As we briefly discussed in the introduction, stars with a mass large enough to collapse into a BH can be modeled to have a PL mass distribution as the progenitor stars' initial mass does \cite{Kroupa:2000iv}. 
This justifies testing the mass distribution of BHs detected by LVK using a PL distribution (see also \cite{PhysRevX.13.011048}). 
We note that we do not claim that first generation black holes that form binaries have a mass distribution that strictly follows a PL distribution, just that a PL distribution is a reasonable approximation to the mass distribution of first generation BHs. 
Following \cite{Bouhaddouti:2024ena}, we consider two parameters for this model: $\alpha$, with $-\alpha$ the exponent of the $m_1$ distribution, $dN/d m_{1} \propto m_{1}^{-\alpha}$ and $\beta$, the exponent of the ratio $q=m_2/m_1$ distribution ($dN/dq \propto q^{\beta}$).
We fit the values of $\alpha$ and $\beta$ to the LVK observations. 

A better approximation of the mass distribution of first generation BHs is a PL with an exponential cutoff. 
Such a distribution accounts for the mass distribution of the progenitors of first generation BHs, but also accounts for the pair instability mass gap \cite{Woosley:2021xba}, which results from the progenitor star having a mass large enough to start producing electron positron pairs during the core collapse, which releases enough energy to cause a supernova with no BH formation at the center. The lower end of the mass gap, while still debated, is known to be at least $40M_\odot$ \cite{Woosley:2021xba}. We choose the exponential cutoff at  $40M_\odot$. Thus, the PL with an exponential cutoff mass distribution of $m_1$ is taken to be, 
\begin{equation}
    \frac{dN}{dm_1} \propto m_1^{-\alpha}e^{(-m_1/40M_{\odot})}.
\end{equation}
For $q$ we use a simple power-law distribution with an exponent $\beta$. We refer to this option as ``PLe''. 

We test both the PL distribution and the PLe distribution for $m_{1}$ and model the redshift distribution for the first generation BH binaries' comoving merger rate to scale as $R(z) \propto (1+z)^{\kappa}$ following Refs.~\cite{ligo_scientific_collaboration_and_virgo_2023_8177023,LIGOScientific:2025pvj}. We take $\kappa = 2.7$, which is consistent with the star formation rate \cite{Y_ksel_2008, Madau_2014}.

\subsection{Modeling the rates of hierarchical merger binaries}
\label{subsec:hierarchical_mergers}

\begin{figure}[h!]
\includegraphics[keepaspectratio,width=0.989\linewidth]{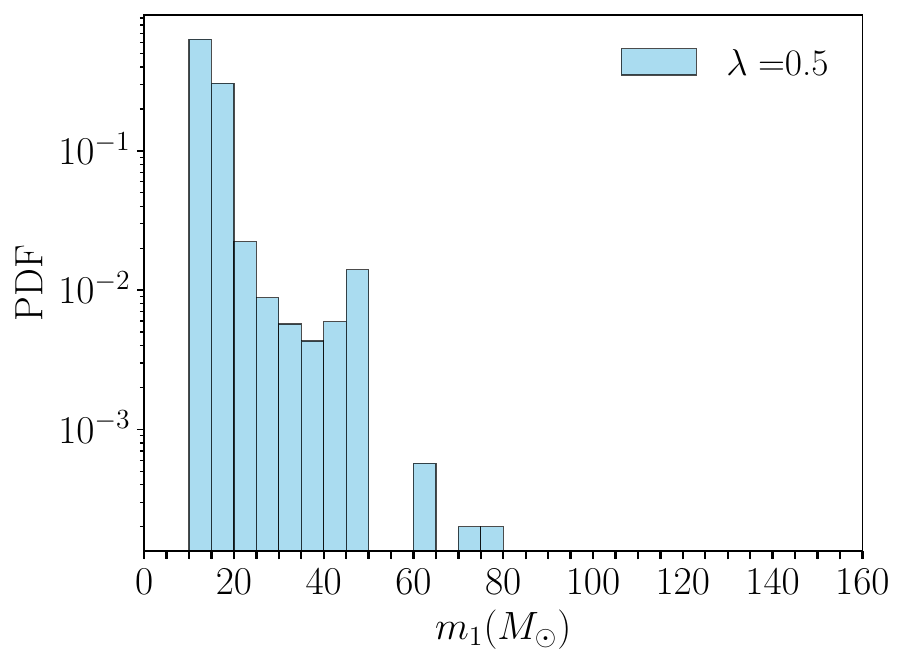} \\
\includegraphics[keepaspectratio,width=0.989\linewidth]{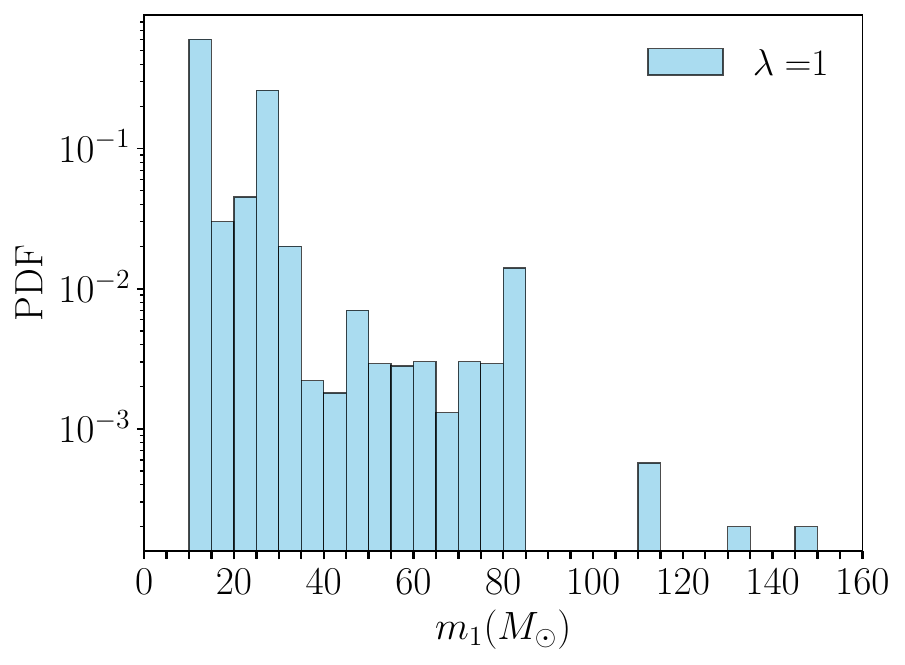} \\    
\includegraphics[keepaspectratio,width=0.989\linewidth]{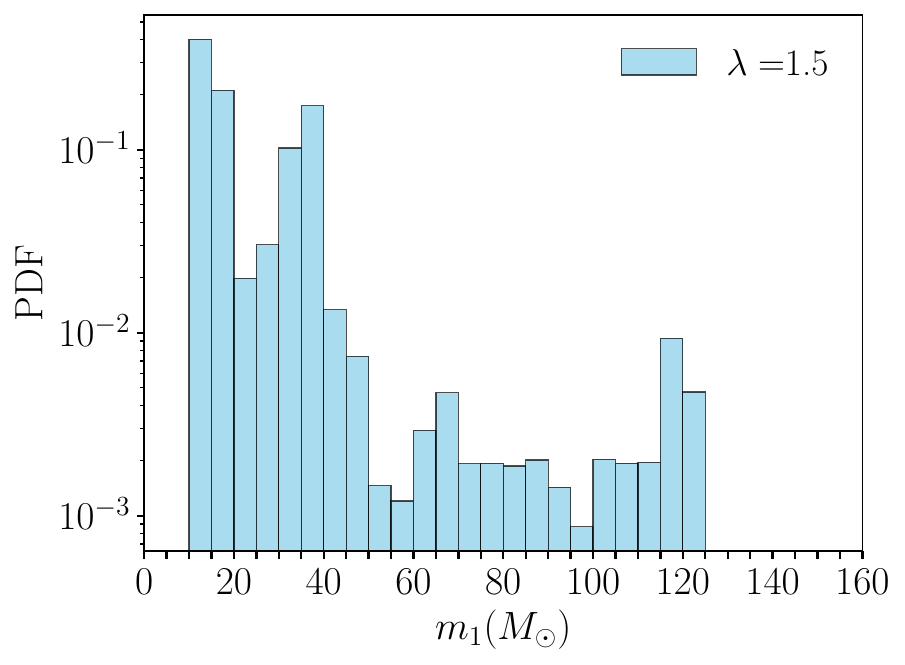}
\caption{PDF of primary mass of BHs that are at least second generation BHs. The top histogram is for $\lambda=0.5$, the middle one for $\lambda=1$ i.e the PDF obtained from \cite{Ye:2025ano}, and the bottom histogram is for $\lambda=1.5$.}
\label{fig:hm_hists}
\end{figure}

By hierarchical merger binaries HM, we refer to mergers of BH binaries where at least one of the BHs is the result of a previous binary BH merger. We consider the $m_1$ and $q$ mass distributions from \cite{Ye:2025ano}, specifically the histograms representing PDFs of higher than first generation BHs in figures 4 and 6 of Ref.~\cite{Ye:2025ano}.
These histograms are the result of N-body simulations of binary evolution and mergers inside dense stellar clusters. 
To allow some modeling flexibility for HM binaries, we introduce an additional parameter $\lambda$ to squeeze or stretch the $m_1$ distribution. In Fig.~\ref{fig:hm_hists}, we show how different values of $\lambda$ affect the  expected $m_1$ distribution from HM binaries. In our tests, parameter $\lambda$ takes values from 0.5 to 1.5 with increments of 0.1. 

From Ref.~\cite{Ye:2025ano}, we also adopt the HM binaries' comoving merger rate scaling with redshift as,
\begin{equation}
    R_{\text{HM}} \propto \frac{(1+z)^{2.6}}{1+[(1+z)/(3.2)]^{6.2}} \, .
\end{equation}

\subsection{PBH merger rates}
\label{subsec:PBH_merger_rates}

PBH binaries can be classified based on their formation time into early binaries and late binaries 
(for a detailed review, see \cite{Raidal:2024bmm}). 
Early binaries form before the matter–radiation equality \cite{Sasaki:2016jop, Ali-Haimoud:2017rtz}. 
The evolution of their orbital properties follows two main pathways: (i) unperturbed binaries, which remain 
outside dark matter halos and just evolve via gravitational-wave emission, and (ii) binaries that stay unperturbed 
initially, but at some point fall inside halos, where  some of them will undergo dynamical effects such as binary–single interactions 
with other PBH binaries or single PBHs. Instead, late binaries form through dynamical  gravitational-wave direct capture 
of single PBHs inside dark matter halos~\cite{Bird:2016dcv}, creating highly eccentric binaries that
merge soon after~\cite{Cholis:2016kqi}. Thus, the combined total comoving merger rate of PBH binaries at redshift can be written as,
\begin{equation}\label{Rtotal}
\begin{split}
R_{\rm total}(z) &= f_{\rm PBH}^2 \Big[ R_{\rm captures}(z) +  2\, f_{\rm PBH\,binaries}\times \\&\big( R_{\rm unperturbed}(z) + R_{\rm binary-single}(z) \big) \Big],
\end{split}
\end{equation}
where $R_{\textrm{unperturbed}}$ is the  comoving merger rate from unperturbed early binaries merging outside dark matter halos, 
$R_{\textrm{binary-single}}$ is the merger rate of early binaries further evolved through binary-single interactions inside halos, 
and $R_{\textrm {capture}}$ is the rate of late binaries formed via gravitational-wave induced captures. $f_{\rm PBH\,binaries}$ denotes the fraction of PBHs initially bound in binaries.
For the allowed PBH merger rate calculation, we take as a starting point $f_{\rm PBH}=1$, 
which is then constrained by the LVK observations.

We use the recent calculations of Refs.~\cite{Aljaf:2024fru, Aljaf:2025dta}, where each of these rates has been  re-evaluated, and the relevant modeling uncertainties have been quantified. 
In those works, a large sample of early binaries is simulated forming at $z=3200$, with their initial semi-major axes and eccentricities $(a,e)$ drawn from the joint PDF evaluated by~\cite{Sasaki:2016jop, Ali-Haimoud:2017rtz, Kavanagh:2018ggo} \footnote{These binaries are evolved to $z=0$ by combining two approaches: unperturbed binaries that remain outside dark matter halos evolve purely via gravitational wave emission following the orbit-averaged Peters equations \cite{1964PhRv..136.1224P}, while binaries that fall into halos are further evolved, including dynamical binary-single interactions (perturbed binaries).}. 
We take that half of PBHs are in binaries and track the fraction of dark matter that stays in the intergalactic space, i.e., does not enter the Virial radius of dark matter halos at any time. 
Our simulations start near the formation of PBH binaries. At this stage, only a small fraction of dark matter is in halos. At $z=12$, this fraction is just a few percent. However, by today, approximately 90\% of all dark matter PBHs are inside halos \cite{Murray:2013qza}.

\begin{figure}
    \centering
    \includegraphics[keepaspectratio,width=0.99\linewidth]{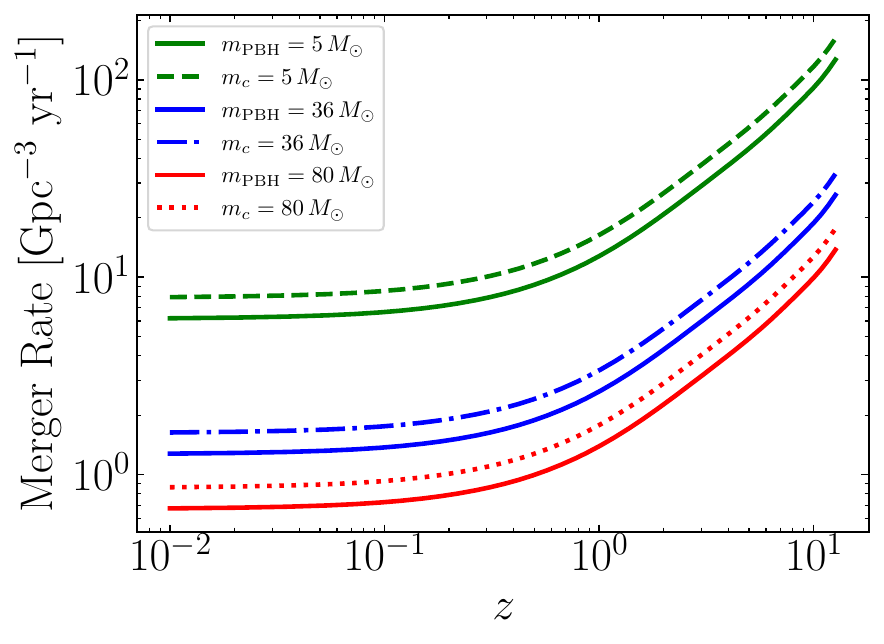}    
    \vspace{-0.2in}
    \caption{The  comoving merger rate of PBHs  binaries as a function of redshift $z$ for monochromatic and for lognormal mass distributions described by Eq.\eqref{eq:MassPDF_lognormal} with $\sigma=0.6$. The rate is calculated as the sum of three  merger channels:  unperturbed binaries,  binary-single interactions within halos, and two body  captures using Eq.\eqref{Rtotal}. In all cases, we assume $f_{\rm PBH}=3\times 10^{-3}$ and a binary fraction of $f_{\rm PBH\,binaries}=0.5$.}
    \label{fig:Total_rates:channels}
\end{figure}

PBH binaries that fall inside halos will experience dynamical binary-single interactions that can lead to 
mergers, hardening, softening, or even ionization (binaries breaking up)~\cite{Heggie:1975tg, 1987Binney, Quinlan:1996vp, Sesana:2006xw}. Hard binaries become more tightly bound, soft binaries are widened, and sufficiently strong encounters can break a binary \cite{Heggie:1975tg}. 
We follow the methodology of Ref.~\cite{Aljaf:2025dta} and account for all these effects on the evolution of PBH binaries.
The rate of binary-single interactions depends on the local density and  dispersion velocity (see also discussion 
in \cite{Aljaf:2024fru}). For this reason, the merger rates are evaluated for dark matter halos of masses 
$10^{4} - 10^{15} \, M_{\odot}$, properly tracking the dark matter halo mass function evolution with time. 
The density of PBHs in each halo follows an NFW profile~\cite{Navarro:1995iw}.
In each simulated dark matter halo, the simulations account for the change of density and dispersion velocity with 
radius and with time within the halo, which affects the environment that the PBH binaries are in 
(see \cite{Aljaf:2024fru, Aljaf:2025dta} for further details). All dark matter halos are simulated to grow in mass with time 
following Ref.~\cite{Correa:2015kia}. The orbital properties of all binaries inside a halo are tracked through 
orbit-averaged evolution equations that can lead to hardening all the way to merger, to softening of the binaries, and even to ionizations \cite{Aljaf:2025dta}.

Finally, for the late binaries formed inside dark matter halos via gravitational-wave two-body direct captures, their rate of merger is determined using the analytic treatment of~\cite{Sasaki:2016jop, Bird:2016dcv, Aljaf:2024fru}. 

We calculate the comoving merger rate using \eqref{Rtotal}, assuming both monochromatic and lognormal PBH mass distributions.
In the monochromatic case, all PBHs have identical source-frame masses 
$m_{1}=m_{2}=m_{\rm PBH}$, while in the lognormal case the mass function is
\begin{equation}
\psi(m)=\frac{1}{\sqrt{2\pi}\,\sigma\, m}
\exp\!\left(-\frac{\ln^2(m/m_c)}{2\sigma^2}\right),
\label{eq:MassPDF_lognormal}
\end{equation}
where $m_c$ denotes the peak mass and $\sigma=0.6$ is the width of the distribution. 
In both cases, the characteristic mass ($m_{\rm PBH}$ or $m_c$) is taken in the range 
$5 M_{\odot} \le m \le 80 M_{\odot}$.

In Fig.~\ref{fig:Total_rates:channels} we present the comoving merger rate as a function of redshift $z$ for three representative masses $m_{\rm PBH}=m_c=5,\ 36,$ and $80\,M_{\odot}$. Solid lines correspond to the monochromatic case, while dashed, dot-dashed, and dotted lines correspond to the lognormal distribution. For these calculations  we assume $f_{\rm PBH}=3\times10^{-3}$ and a binary fraction $f_{\rm PBH\ binaries}=0.5$. We find that for the same characteristic mass the lognormal distribution with $\sigma=0.6$ in Eq.~\eqref{eq:MassPDF_lognormal} provides slightly higher merger rates.

\subsection{Simulating detectable black hole binary mergers}
\label{subsec:SimulatingBinaries}

To simulate BH binaries, we first draw $m_{1}$ masses from the considered population; 
i) first generation BH binaries with a PL distribution for $m_{1}$, ii) first generation BH binaries with a PLe distribution, i.e., PL and an exponential cutoff for $m_{1}$ at $40 \, M_{\odot}$, iii) HM binaries, or iv) PBH binaries. For each population, we use the relevant $q$-distribution and then take the product $m_1\cdot q$ to obtain the $m_2$ mass. 
We also assign a redshift value to each simulated BH binary that is consistent with the merger rate dependency on $z$. 
In this work, for every merging binary, we also assign a time of arrival of the gravitational waves at Earth. 
This arrival time is drawn from a uniform distribution that covers the dates the LIGO runs were occurring. This allows us to take into account all the power spectral densities (PSDs) provided by LIGO \cite{ligo_scientific_collaboration_and_virgo_2023_8177023,ligo_scientific_collaboration_and_virgo_2025_16053484}, to correct in our parameter estimation for the fact that the two LIGO sensitivity curves changed with time. 

As discussed previously, all reported masses have a measurement error. To account for that in our simulations, we first derive the distribution of $m_{1}$ values of the simulated merger events with an SNR $\ge 8$ for specific model parameters describing the simulated population of BBHs. We then convolve that distribution with a Gaussian function. That Gaussian function has a mass dependent width based on the reported errors on the $m_{1}$ and $m_{2}$ values of the LVK events, see Refs.~\cite{Bouhaddouti:2024ena, Bouhaddouti:2025ltb} for more details. The uncertainty on the $m_{2}$ masses is accounted for as the simulated $m_{2}$-distribution is the product of the $m_{1} \cdot q$ distribution, where the $m_{1}$-distribution has already been convolved with the Gaussian described above. 

To compare our models to the data of Fig.~\ref{fig:m_1_2_z_hists}, we calculate the SNR for binaries at chosen values of their relevant modeling parameters, e.g., $\alpha$ and $\beta$ for the PL and the PLe distributions. 
The choice of PSD depends on the time of gravitational wave arrival at Earth from the simulated binary. 
We use the PSD that is closest to the simulated gravitational wave arrival time. 
The SNR equation for each merged BH binary depends on whether one or both LIGO detectors were operational at the time the gravitational waves arrived at Earth. When both LIGO detectors are operational, we use, 
\begin{equation}
    SNR^{2} = \frac{2}{5}  \int_{f_{\textrm{min}}}^{f_{\textrm{max}}} df \frac{h_c(f)^{2}}{(2f)^2} \left(\frac{1}{\text{PSD}_H(f)} +  \frac{1}{\text{PSD}_L(f)} \right).
\end{equation}
The indices $H$ and $L$ refer to the Hanford or Livingston detectors respectively, $f$ is the observed gravitational-wave frequency, $f_{\textrm{min}}$ and $f_{\textrm{max}}$ are the current range of LIGO's interferometers (taken to be 10 Hz to 4000 Hz), $h_{c}(f)$ the observed strain amplitude, PSD($f$) = $h_{n}(f)^2$ ($h_{n}(f)$ the strain noise amplitude).
If only one of the detectors is operational we use instead,
\begin{equation}
   SNR^{2} = \frac{2}{5}\int_{f_{\textrm{min}}}^{f_{\textrm{max}}} df \frac{h_c(f)^{2}}{\text{PSD}(f) \, (2f)^2}.
\end{equation}
The PSD in the above equation is the relevant  power spectral density of the operational detector. For a detailed description of $h_c$ and the equations used for the SNR calculation, we refer the reader to \cite{Bouhaddouti:2024ena} and \cite{Flanagan_1998}.

The total number of simulated merging binaries between redshifts $z_{1}$ and $z_{2}$, is obtained using,
\begin{equation}
    N_{\textrm{merger}}(z_1,z_2)= 4\pi \int_{z_1}^{z_2} dz \frac{c \cdot \chi(z)^2 \cdot R_{\textrm{pop}}(z)}{(1+z)H(z)} \, \times t_{obs},
\end{equation}
where $\chi(z)$ is the comoving distance, $R_{\textrm{pop}}(z)$ the comoving merger rate of the considered population, $H(z)$ the Hubble expansion parameter and $t_{obs}$ the duration of the runs O1 to O4a taking into account the duty cycle of each run. 
By adding all redshift bins, this allows us to generate histograms for $m_{1}$ and $m_{2}$ of the expected detectable BH binaries from any given modeled population, while also being able to keep track of the redshift distribution of those mergers. 
We then compare to the histograms of Fig.~\ref{fig:m_1_2_z_hists}. 
To minimize the statistical noise on the 
predicted histograms of $m_{1}$, $m_{2}$ and $z$,
we  multiply the $R_{\textrm{pop}}(z)$ by some factor (between 10 and 1000 depending on the population), which we keep track of and properly remove in our reported constraints on the merger rates of the various populations.

\subsection{Best fit parameter estimation with \texttt{emcee}}
\label{sebsec:Best_fit}

We use \texttt{emcee} \cite{2013PASP..125..306F}, to find the best fit parameters of the considered models when comparing to the data. This involves defining a log-likelihood and a uniform prior distribution for each considered model parameter within the ranges given in Table~\ref{table:priors}. 
\begin{table}[h]
\centering
\begin{tabular}{| c | c | c|} 
 \hline
  Model & Parameter & Range \\ [0.8ex] 
 \hline
 \multirow{2}{8em}& $\alpha$ & [2.5, 4] \\ 
 {PL} & $\beta$ & [0, 2.5]\\
  & $R_{\textrm{PL}}$ & [0, $2 \times 10^{2}$] $\textrm{Gpc}^{-3} \textrm{yr}^{-1}$ \\
 \hline
 \multirow{3}{8em} &  $\alpha$ & [2, 4] \\ 
 {PL with cutoff (PLe)} & $\beta$ & [0, 2]\\
  & $R_{\textrm{PLe}}$ & [0, $2\times 10^{2}$] $\textrm{Gpc}^{-3} \textrm{yr}^{-1}$\\
  \hline
  \multirow{3}{8em}  & $\lambda$ & [0.5, 1.5] \\ 
 {HM} & $R_{\textrm{HM}}$ & [0, $2 \times 10^{2}$] $\textrm{Gpc}^{-3} \textrm{yr}^{-1}$\\   
 \hline
  \multirow{3}{8em} &  $m_{\textrm{PBH}}$ & [5, 80] $M_{\odot}$ \\ 
 {Mono PBH} & $f_\textrm{PBH, mono}$ & [0, 1]\\ 
 \hline
  \multirow{3}{8em} &  $m_{c}$ & [5, 80] $M_{\odot}$ \\ 
{Lognorm PBH} & $f_\textrm{PBH, lognorm}$ & [0, 1]\\ 
 \hline
\end{tabular}
\caption{Range of uniform priors used for each model. PL refers to our simple power-law model for $m_{1}$, PLe to the model that has a power-law with an exponential cutoff for $m_{1}$, HM to hierarchical mergers, Mono PBH to monochromatic PBHs, and Lognorm PBH to PBHs drawn from a lognormal mass-distribution. The reported rates are at $z=0$ (see text for more details).}
\label{table:priors}
\end{table}

We use the representation of the LVK data from Fig. \ref{fig:m_1_2_z_hists}. We bin the simulated BBHs from the considered model the same way we bin our data. Since the data represents a number of counts per bin, for the log-likelihood function we use is the Poisson log-likelihood,
\begin{equation}
    ln(\mathcal{L})=\Sigma_ik_iln(\mu_i) - \mu_i -ln(k_i!),
    \label{eq:PoissonLikelihood}
\end{equation}
where $i$ refers to the bin $i$: $i\leq15$ refers to the bins of the $m_1$ histogram, $15<i\leq30$ refers to the bins of the $m_2$ histogram and $30<i\leq40$ refers to bins of the $z$ histogram \footnote{We recognize that using the Poisson log-likelihood assumes no correlations between neighboring bins. Depending on the exact amplitude of the gravitational-wave signals the posterior PDFs of each event's properties typically span more than one bins in any of the three parameters. Thus a Poisson likelihood is only an approximation.}. In Eq.~\ref{eq:PoissonLikelihood}, $k_i$ is the number of counts in bin $i$ of the LVK data, while $\mu_i$ is the number of counts in bin $i$ of the simulated BBHs from the considered model: $\mu_i = \textit{bin counts} \,(\textrm{parameters, $i$)}$.
When we test to the data the PL or PLe, their parameters are $\alpha$, $\beta$, i.e. the exponents of the $m_1$ and $q$ distribution, respectively and also the merger rate $R_{\textrm{PL}}$ (or $R_{\textrm{PLe}}$).
When HMs are considered as an additional population in the fit, the parameter $\lambda$ is included, which describes the squeezing/stretching of the HM $m_{1}$-mass distribution and the merger rates of the HM population $R_{\textrm{HM}}$. Again, we use flat priors for all our model parameters within the probed volume, shown in Table~\ref{table:priors}. The quoted rates are the comoving rates normalized at $z=0$. 

When running \texttt{emcee} to maximize log-likelihood, we use 32 walkers for 50000 iterations, disregarding the first 5000 steps. To get the posterior probabilities, we use the \texttt{`sampler.get\_chain'} function after using the \texttt{`sampler.run\_mcmc'} function. 
We place the walkers initially at $\alpha = 3.5$, $\beta=1.1$ that is the best fit value from Ref.~\cite{KAGRA:2021duu}. 

The LVK collaboration gives the estimated BBH rate  to be $\simeq 20 \, \textrm{Gpc}^{-3} \textrm{yr}^{-1}$. In our priors to allow for maximum modeling freedom, we allow for the rates to go as high as $200 \, \textrm{Gpc}^{-3} \textrm{yr}^{-1}$.
For the HM, we set $\lambda = 1$, which corresponds to the original proposed distribution of Ref.~\cite{Ye:2025ano}, and for the PBHs, our walkers start from  $m_{\textrm{PBH}}=35$, that was the best fit value from O1-O3 LVK observations \cite{Bouhaddouti:2025ltb}. 
Our walkers have initial values for the normalization of the merger rates at $z=0$ to be $20 \, \textrm{Gpc}^{-3} \textrm{yr}^{-1}$ for each of the PL, the PLe, and the HM populations and $f_\textrm{PBH} = 10^{-3}$ for the PBHs. The normalization rates for the first three populations are chosen to be close to the reported results from Ref.~\cite{LIGOScientific:2025pvj}, while the initial walkers' value on $f_\textrm{PBH}$ is at the same order of magnitude of Ref.~\cite{Bouhaddouti:2025ltb}. 
We have also tested the sensitivity of our final results to the initial position of the walkers. 
We run \texttt{emcee} with random initial/starting positions for our walkers within the ranges of our priors of Table~\ref{table:priors}, with the same number of walkers, steps, and discarded steps as previously described. 
Our results are practically unaffected by the choice on the starting position of our walkers.     

\section{Results} 
\label{sec:results}

\begin{figure*}[ht]
    %\centering
    \includegraphics[keepaspectratio,width=0.85\linewidth]{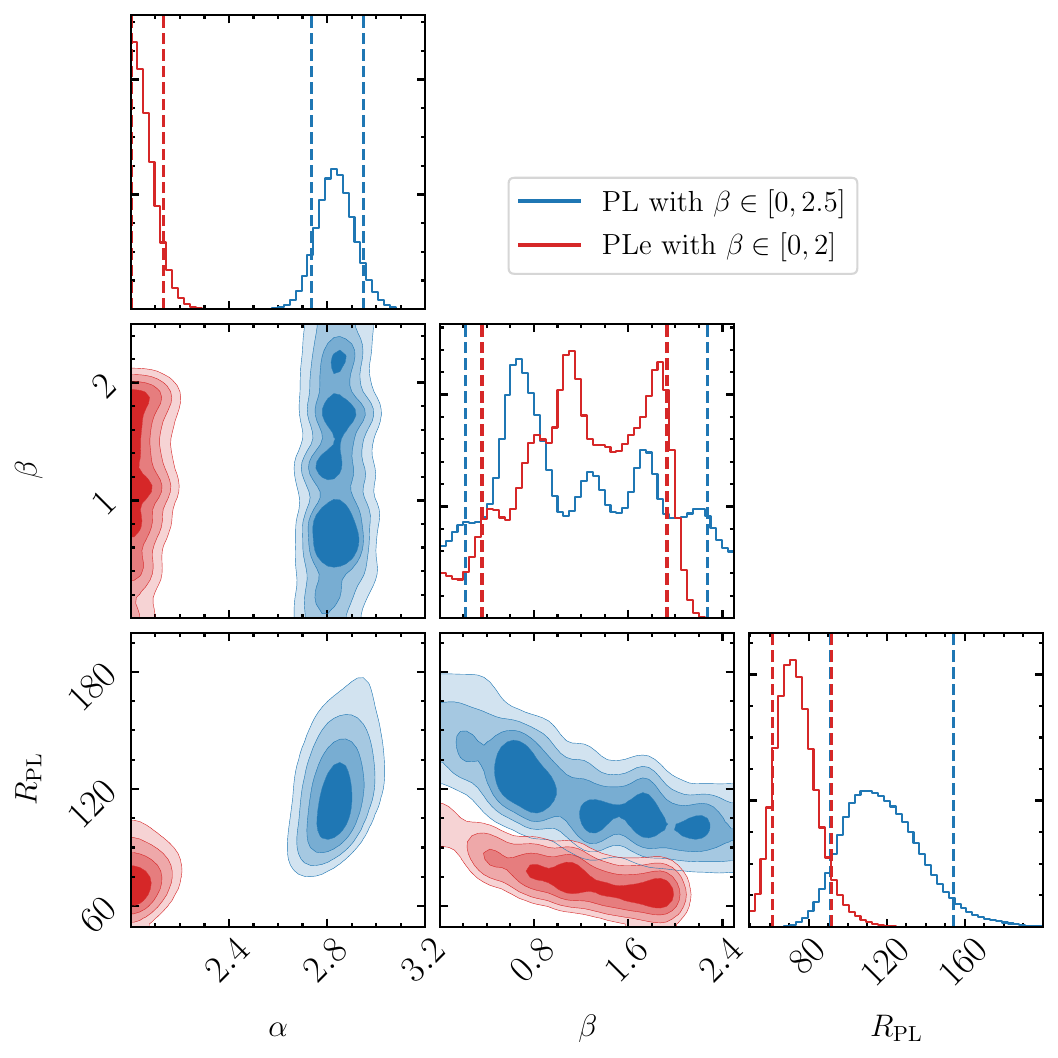}
    \vspace{-0.1in}
    \caption{Corner plot representing the fit parameter values for a simple power-law model (blue lines and contours) and the fit parameter values for a power-law with an exponential cutoff (red lines and contours). The contour lines represent the $34\%$, $68\%$, $85\%$ and $95\%$ credible intervals. We also show in the dashed lines the 90$\%$ credible interval ranges for each of the parameters values are given in Table~\ref{table:posteriors_neg_beta}. We universally use the symbol $R_{\textrm{PL}}$ to describe the comoving merger rate parameter in $\textrm{Gpc}^{-3}\,\textrm{yr}^{-1}$ at $z=0$  (see text for more details).}
    \label{fig:PL_corner_plot}
\end{figure*}

In Fig.~\ref{fig:PL_corner_plot}, we show the results of the fit for parameters of $\alpha$, $\beta$, and the associated merger rate of the PL and the PLe models.
For the case of the simple power-law model (blue regions, lines and contours) for $m_{1}$, we find $\alpha$ to be $2.84^{+0.11}_{-0.10}$, within $90\%$ credible interval (CI). The central values represent the median value from the posterior PDF. We also find $\beta = 1.11^{+1.17}_{-0.89}$ at $90\%$ CI and the associated comoving merger rate at $z=0$ to be $R_{\textrm{PL}} = 117^{+37}_{-25}$ $\textrm{Gpc}^{-3}\textrm{yr}^{-1}$ at $90\%$ CI.  Instead, for the case where the PLe model is tested we find $\alpha = 2.04^{+0.09}_{-0.04}$, $\beta = 1.19^{+0.74}_{-0.83}$ and 
$R_{\textrm{PLe}} = 73^{+19}_{-11}$ $\textrm{Gpc}^{-3}\textrm{yr}^{-1}$ at $90\%$ CI (red regions, lines and contours).  
We also give these relevant ranges in Table~\ref{table:posteriors_neg_beta}, for ease of comparison. 

\begin{table*}[ht]
\centering
\begin{tabular}{| c | c | c | c | c | c | c | c | c | c | } 
 \hline
  Model & $\alpha$ & $\beta$ & $R_{\textrm{PL}}$ & $\lambda$ & $R_{\textrm{HM}}$ & $m_{\textrm{PBH}}$ & $f_{\textrm{PBH}}$ & $- 2 \Delta ln\mathcal{L}$ & $ln\textrm{BF}$\\ [0.8ex] 
   & &  & or $R_{\textrm{PLe}}$ & & & or $m_{c}$ &  ($\times 10^{-3}$) & & \\ 
 \hline
 PL & $2.84^{+0.11}_{-0.10}$ & $1.11^{+1.17}_{-0.89}$ & $117^{+37}_{-25}$ & - & - & - & - & 0 & 0 \\
 \hline
 PLe & $2.04^{+0.09}_{-0.04}$ & $1.19^{+0.74}_{-0.83}$ & $73^{+19}_{-11}$ & - & - & - & - & -40 & +21 \\
 \hline
 PL $\&$ HM & $3.45^{+0.44}_{-0.60}$ & $0.76^{+1.49}_{-0.70}$ & $61^{+52}_{-58}$ & $1.46^{+0.03}_{-0.43}$ & $7.4^{+11.1}_{-1.4}$ & - & - & -144 & +66 \\
 \hline
 PLe $\&$ HM & $2.51^{+0.38}_{-0.35}$ & $0.66^{+1.14}_{-0.61}$ & $54^{+29}_{-23}$ & $1.48^{+0.02}_{-0.13}$ & $5.9^{+2.0}_{-1.2}$ & - & - & -150 & +71 \\
 \hline
 PL $\&$ HM $\&$ Mono PBH  &  $3.54^{+0.41}_{-0.83}$ & $0.95^{+1.36}_{-0.88}$ & $15^{+49}_{-14}$ & $1.14^{+0.03}_{-0.42}$ & $13.2^{+5.7}_{-5.1}$ & $38.7^{+3.4}_{-2.3}$ & $3.39^{+0.49}_{-0.52}$ & -185 & +79 \\
 \hline
 PLe $\&$ HM $\&$ Mono PBH  &   $2.98^{+0.88}_{-0.75}$ & $0.83^{+1.03}_{-0.76}$ & $21^{+116}_{-19}$ & $1.15^{+0.01}_{-0.39}$ & $14.6^{+4.8}_{-6.4}$ & $39.0^{+3.4}_{-2.3}$ & $3.35^{+0.50}_{-0.53}$& -186 & +80 \\
 
 \hline
 PL $\&$ HM $\&$ Lognorm PBH  &  $3.53^{+0.42}_{-0.92}$ & $0.98^{+1.35}_{-0.91}$ & $8.4^{+45.6}_{-7.9}$ & $1.20^{+0.02}_{-0.66}$ & $10.8^{+4.4}_{-5.6}$ & $32.7^{+5.5}_{-3.5}$ & $6.72^{+0.85}_{-1.18}$ & -216 & +96 \\
 \hline
 PLe $\&$ HM $\&$ Lognorm PBH  &  $2.98^{+0.89}_{-0.56}$ & $0.85^{+1.02}_{-0.78}$ & $23^{+136}_{-22}$ & $1.20^{+0.02}_{-0.58}$ & $11.4^{+4.2}_{-5.6}$ & $32.5^{+5.1}_{-3.4}$ & $6.76^{+0.83}_{-1.07}$ & -216 & +97 \\
 \hline
\end{tabular}
\caption{The fit ranges for the tested models. We present here the results for the prior assumption that $\beta \ge 0$ (we give the results for alternative priors on $\beta$ and $R_{\textrm{PL}}$ or $R_{\textrm{PLe}}$ in Appendix~\ref{app:Alt_priors}).
The central values give the median value from the posterior PDF. The upper and lower values define the $90\%$ credible interval range that are also represented in the relevant corner plots as dashed lines.
Parameters $\alpha$, $\beta$, $\lambda$ and $f_{\textrm{PBH}}$ are without units, $R_{\textrm{PL}}$ and $R_{\textrm{HM}}$ are in $\textrm{Gpc}^{-3} \textrm{yr}^{-1}$ and $m_{\textrm{PBH}}$ or $m_{c}$ are in $M_{\odot}$.
The second to last column gives the $-2\Delta ln \mathcal{L}$ between any population synthesis model ``$X$'' and the simple PL model, i.e. $\Lambda^{X}_{PL}$. The last column gives in turn the logarithm of the Bayes factor $ln\textrm{BF}^{X}_{\textrm{PL}}$ in comparing model ``$X$'' to the PL model. In evaluating the range of values on $f_{\textrm{PBH}}$, we have assumed $f_{\textrm{PBH binaries}}= 0.5$.}
\label{table:posteriors_pos_beta}
\end{table*}
While both the PL and PLe are simplified models and we have clear evidence already from the O1-O3 runs for an additional ``peak'' around 30-40 $M_{\odot}$ in $m_{1}$. 
Still these models provide a basis to compare with the more complex models of an added HM population or adding both a HM BH population and a PBH population, that we discuss in the following paragraphs. 
Our results for the posterior ranges of $\alpha$ and $\beta$ of the PL model, overlap with the results of Ref.~\cite{KAGRA:2021duu}, that suggested $\alpha= 3.5^{+0.6}_{-0.6}$ and $\beta = 1.1^{+1.7}_{-1.3}$ for the power-law with a peak model. 
Furthermore, Ref.~\cite{LIGOScientific:2025pvj}, using the same sample of BBH observations suggested a broken power-law distribution with an index $\alpha_{1} = 1.7^{+1.2}_{-1.8}$ for $m_{1} \lesssim 35 \, M_{\odot}$ and a different index $\alpha_{1} = 4.5^{+1.6}_{-1.3}$ for the higher masses. 
Our PLe model clearly prefers lower values of $\alpha$ than the simple PL model (harder spectrum for $m_{1}$). 
As this model mostly fits the population of BBHs with $m_{1} < 40 \, M_{\odot}$, it is not surprising that our derived range is in agreement with the range of the $\alpha_{1}$ index of Ref.~\cite{LIGOScientific:2025pvj}.
We find that in both the PL and the PLe models the $\beta$ index is poorly constrained as has been the case also for the LVK collaboration results \cite{KAGRA:2021duu, LIGOScientific:2025pvj}. 
There is still a great level of degeneracy between the $\alpha$ and $\beta$ indices, with a small change in the value of $\alpha$ causing a large change in the values of $\beta$. 
This can be clearly seen in the contours of  Fig.~\ref{fig:PL_corner_plot}, for both PL and PLe models \footnote{There is a small amount of smoothing in plotting the 2D projection of the posterior PDF that explains why in the $\beta$ vs $R_{\textrm{PL}}$ panel, for the PLe model the red contours extend marginally beyond the $\beta = 2$ value.}.   
We note that the small peaks in the posterior PDF of the $\beta$ parameter more evident in the case of the PLe population are a result of us using the binned data of Fig.~\ref{fig:m_1_2_z_hists}. However, those peaks are at the $34\%$ CI range.  
Our results on the merger rates of the PL and PLe population somewhat higher compared to those of 
Ref.~\cite{LIGOScientific:2025pvj}, that gets the binary BH merger rate to be between 14 and 
26 $\textrm{Gpc}^{-3} \textrm{yr}^{-1}$ at $90\%$ CI. 
As a note the PLe gives a lower and better constrained merger rate to the PL one.
In our fits higher rates are allowed due to the lower values of the $\beta$ parameter that we find to be still allowed. 
A smaller value for $\beta$ allows for many BBHs that would go undetected by LVK. In Appendix~\ref{app:Alt_priors} we show results for alternative (wider) priors on $\beta$ and $R_{\textrm{PL}}$ or $R_{\textrm{PLe}}$.

To compare between population models we first calculate their Bayesian evidence $\mathcal{Z_{\textrm{A}}}$,
\begin{equation}
    \mathcal{Z_{\textrm{A}}}=\int d\theta_{\textrm{A}} \mathcal{L}(D|\theta_{\textrm{A}})\pi(\theta_{\textrm{A}}). 
    \label{eq:Bayesian_evidence}
\end{equation}
The index A, here represents the model that we test: i.e. i) PL for the power-law model, ii) PLe for the power-law with exponential cutoff, iii) PL $\&$ HM and iv) PLe $\&$ HM when considering hierarchical mergers to the PL and the PLe models respectively, v) PL $\&$ HM $\&$ Mono PBH and vi) PLe $\&$ HM $\&$ Mono PBH when also adding a population of monochromatic PBHs to models (iii) and (iv) respectively. We also test instead the case where the PBHs follow a lognormal mass distribution as described in Sec.~\ref{sec:method}.  Those are cases vii) PL $\&$ HM $\&$ Lognorm PBH and viii) PLe $\&$ HM $\&$ Lognorm PBH. 
The prior distribution on the physical parameters $\theta_{\textrm{A}}$, are represented as $\pi(\theta_{\textrm{A}})$ and are described in Table~\ref{table:priors}. 
We use \texttt{emcee} to get the likelihood function $\mathcal{L}(D|\theta_{\textrm{A}})$ of the data $D$ of Fig.~\ref{fig:m_1_2_z_hists}, given the parameters $\theta_{\textrm{A}}$.
We present different assumptions on the priors of these models in Appendix~\ref{app:Alt_priors}.
To obtain the evidence for each model we use version 2.2.1 of \texttt{emcee} \cite{2013PASP..125..306F}, specifically the \texttt{`sampler.thermodynamic\_integration\_log\_evidence'} function after using \texttt{`PTSampler'} \cite{2021ascl.soft01006V} with a number of temperatures equal to $20$, the same number of steps and walkers as described in \ref{sebsec:Best_fit} and the number of dimensions corresponding to the model for which the evidence is calculated. It is important to note that we used version 2.2.1 of \texttt{emcee} only for the estimation of the evidence for the models we consider in this work. The posterior samples, likelihoods are obtained with the most recent version of \texttt{emcee} available during this work (3.1.6). 

The population synthesis models that we test have different number of parameters describing them and thus different prior distributions $\pi(\theta_{\textrm{A}})$ expand to wider parameter spaces. Thus, the Bayes factor when comparing model $A$ to $B$, 
\begin{equation}
    \textrm{BF}^{A}_{B}=\frac{\mathcal{Z_{\textrm{A}}}}{\mathcal{Z_{\textrm{B}}}}, 
    \label{eq:Bayes_factor}
\end{equation}
formally is not enough to properly compare between the alternative models for the population synthesis of the LVK data $D$ that we test. 
We can use that criterion to properly compare between models of same complexity in the parameters probed as for instance the PL and the PLe for which we use the same prior distributions for the same parameters: $\alpha$, $\beta$ and merger rate of the given population evaluated at $z=0$ and having the same parametric dependence to redshift of $\propto (1+z)^{2.7}$.
We can also use the Bayes factor when comparing models (iii) to (iv) i.e. PL or PLe $\&$ HM and between models (v), (vi), (vii) and (viii) i.e. PL or PLe $\&$ HM $\&$ some PBH population. 

We report the $ln\textrm{BF}^{A}_{B} = ln\mathcal{Z_{\textrm{A}}} - ln\mathcal{Z_{\textrm{B}}}$ and also the likelihood ratio,  
\begin{equation}
\Lambda^{A}_{B}= -2 ln\left[ \frac{\mathcal{L_{\textrm{A}}^{\textrm{max}}}}{\mathcal{L_{\textrm{B}}^{\textrm{max}}}}\right]= -2 (ln\mathcal{L_{\textrm{A}}^{\textrm{max}}} - ln\mathcal{L_{\textrm{B}}^{\textrm{max}}}), 
    \label{eq:likelihood_ratio}
\end{equation}
where $\mathcal{L_{\textrm{A}}^{\textrm{max}}}$ and $\mathcal{L_{\textrm{B}}^{\textrm{max}}}$ are the maximum likelihoods for each model, achieved for their relevant best-fit parameter values. We also refer to this as $-2\Delta ln\mathcal{L}$.

Comparing PL to PLe models we find a preference for the PLe model by $ln\textrm{BF}^{\textrm{PLe}}_{\textrm{PL}} = 21$, which is considered a strong evidence in favor of the PLe model (given also in Table~\ref{table:posteriors_pos_beta}) \footnote{For each of the population synthesis models, its individual $ln\mathcal{Z}$ comes with an error. The reported errors from \texttt{emcee} on those $ln\mathcal{Z}$ are $\simeq 5$.}. We find that the PLe model while it predicts fewer high-mass BBHs it explains better the lower mass end of the LVK observations. Moreover, we find that the preference for a PLe model to a PL even when adding HM or HM and PBHs is persistent as we explain in the following paragraphs.
Furthermore, we find a likelihood ratio between PLe and PL to be $\Lambda^{\textrm{PLe}}_{\textrm{PL}}= -40 $, i.e. in favor of the PLe model (see also Table~\ref{table:posteriors_pos_beta}). 

\begin{figure*}[ht]
    \centering
    \includegraphics[keepaspectratio,width=0.99\linewidth] {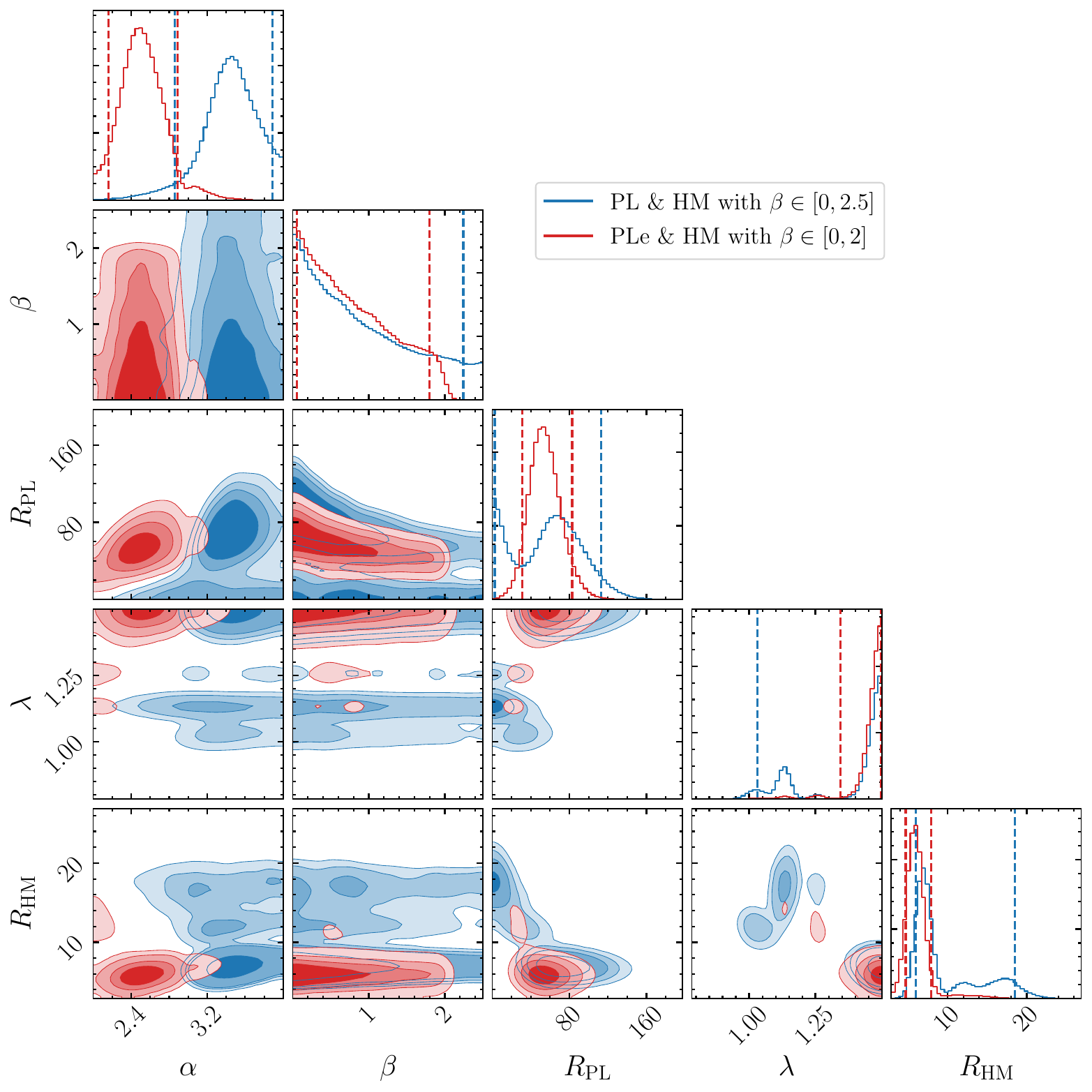}
    \vspace{-0.1in}
    \caption{Similarly to Fig.~\ref{fig:PL_corner_plot}, corner plot representing the fit parameter values for the PL $\&$ HM model (blue color) and the PLe $\&$ HM model (red color). Contours again represent the $34\%$, $68\%$, $85\%$ and $95\%$ CIs and the vertical dashed lines give the 90$\%$ CI parameter value ranges. Parameters $\alpha$, $\beta$ and $\lambda$ are without units, while $R_{\textrm{PL}}$ and $R_{\textrm{HM}}$ are in $\textrm{Gpc}^{-3}\textrm{yr}^{-1}$.}
    \label{fig:PLe_HM_corner_plot}
\end{figure*}

In Fig.~\ref{fig:PLe_HM_corner_plot}, using PL and the PLe populations for the first generation BHs, we add a HM population and probe the five-parameter space (see Table~\ref{table:priors}). We find that the PL and PLe components of these models give the dominant contribution to the LVK mergers but more importantly increase the inferred physical rate, with the PL or PLe merger rates ($R_{\textrm{PL}}$ in the panels of Fig.~\ref{fig:PLe_HM_corner_plot}) being up to a factor of $20$ higher than the HM merger rates $R_{\textrm{HM}}$. Adding an HM population allows in our fits for the PL (PLe) populations to adjust better their model parameters to fit the lower LVK masses of Fig.~\ref{fig:m_1_2_z_hists}.
Even for these models as was the case for the simple PL and PLe models, the $\alpha$ parameter for the PLe component has lower (less steep) power-law values of $2.51^{+0.38}_{-0.35}$ compared to the PL component ($3.45^{+0.44}_{-0.60}$). The $\beta$ parameter for these components is effectively unconstrained but with a preference for $\beta < 1$ which explains also the high values we get for $R_{\textrm{PL}}$. In Appendix~\ref{app:Alt_priors}, we show results with a prior of $- 1 \leq \beta$ where the derived ranges of $R_{\textrm{PL}}$ rates can be significantly larger.

A main result form Fig.~\ref{fig:PLe_HM_corner_plot}, is also that the $\lambda$ parameter that describes the stretching of the $m_{1}$ HM second generation+ population has a clear preference for $\lambda> 1$ with posterior getting values of $\lambda \simeq 1.5$ which is the end of the probed prior and allows for the HM population to expand to higher masses than originally predicted by Ref.~\cite{Ye:2025ano}. This may suggest the existence of third and even higher generation BHs in the LVK sample (see e.g. Ref.~\cite{Mai:2025jmk, 2026arXiv260204176N}). 

\begin{figure}
    \includegraphics[keepaspectratio,width=0.99\linewidth]{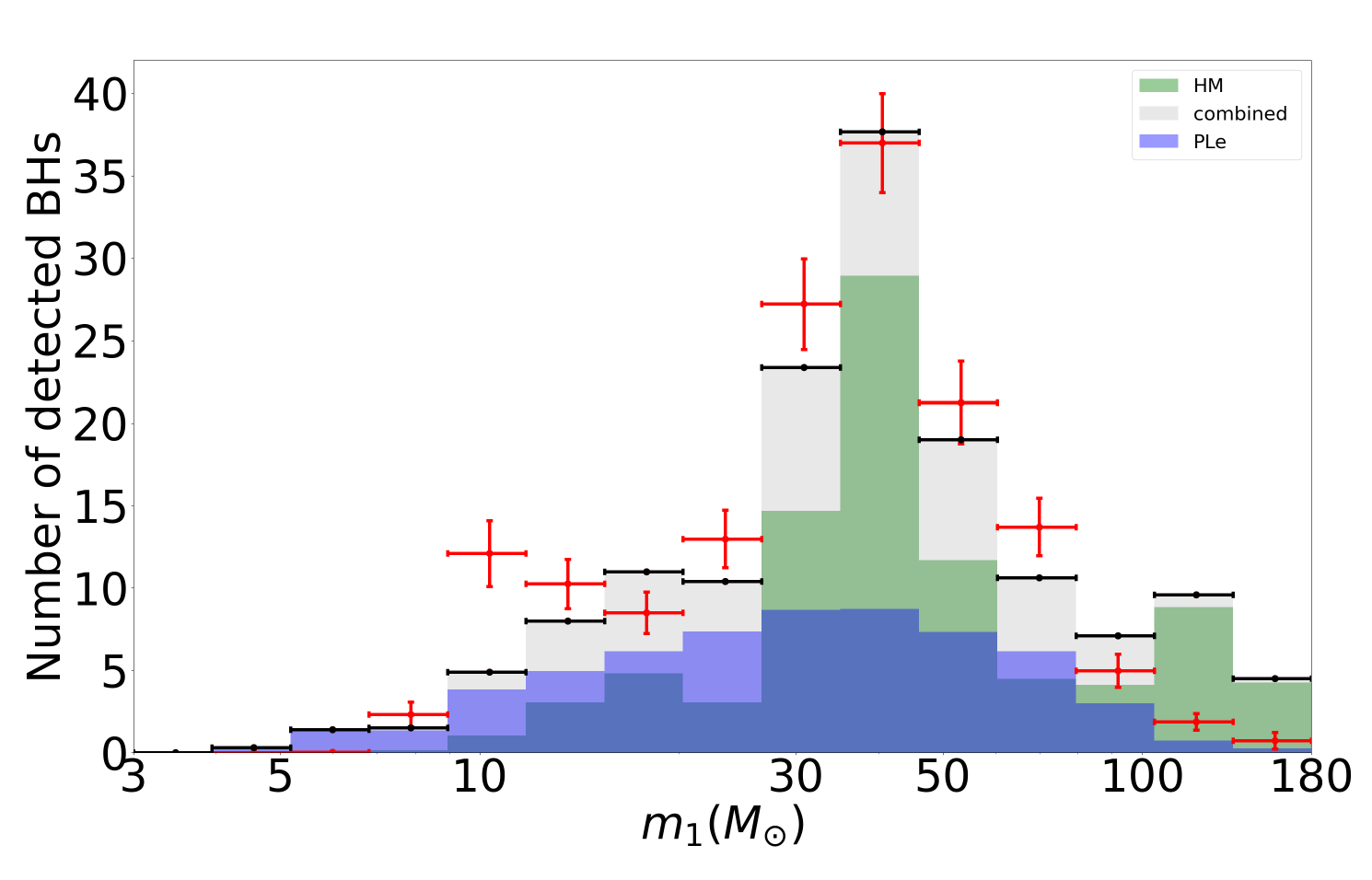}  
    \vspace{-0.3in}
    \caption{The $m_{1}$ simulated histogram from a  PLe $\&$ HM population synthesis model. In the blue histogram we show the $m_{1}$ BHs from the PLe component, while in the green histogram that clearly peaks at 40 $M_{\odot}$, we show the HM contribution. The combined histogram is given by the gray histogram. The LVK observations are shown by the red (Poisson) error-bars. The comoving merger rate of the PLe component $R_{\textrm{PLe}}$ is more than 10 times larger than the rate of the HM component $R_{\textrm{HM}}$. However, the populations contribute about the same number of detectable events.}
    \label{fig:M1_PLeandHM_histogram}
\end{figure}

We find that adding a HM component is preferred to a pure PLe model by a  Bayes factor $\textrm{BF}^{\textrm{PLe \& HM}}_{\textrm{PLe}} = 50$ and by a likelihood ratio of $\Lambda^{\textrm{PLe \& HM}}_{\textrm{PLe}}= -110$.
This clear preference for a HM component is also seen when using instead of the PLe model for the first generation BHs the simple PL model. The relevant quantities are 
$\textrm{BF}^{\textrm{PL \& HM}}_{\textrm{PL}} = 66$ and $\Lambda^{\textrm{PL \& HM}}_{\textrm{PL}}= -144$.
The clear evidence for a HM component is due to the significant number of detections of $m_{1} \ge 50 M_{\odot}$, in combination with a few events with $m_{2} \ge 40 M_{\odot}$. 
In Fig.~\ref{fig:M1_PLeandHM_histogram}, we show for the PLe $\&$ HM population synthesis model a histogram of $m_{1}$ and how this HM component explains the high-mass detections. 
In that figure, while the rate for the PLe component $R_{\textrm{PLe}}$ (blue histogram) is more than 10 times larger than the rate of the HM component $R_{\textrm{HM}}$, the HM component (green histogram peaking at $m_{1} \simeq 40 \, M_{\odot}$) still contributes a little more than half the simulated events. Moreover, we note that while we simulated a power-law with an exponential cut off at $40 \, M_{\odot}$ for the first generation BHs, in the detectable sample there are several BHs with $m_{1} \geq 50 \, M_{\odot}$. This is not from some peculiar choice of $\alpha$ or $\beta$ for the PLe population (we used in that figure $\alpha = 2.3$ and $\beta = 0.1$). 
Both the prominent contribution of the HM component and the fact that there are still several BHs from the PLe component at masses beyond the cut off are a result of LVK's higher sensitivity on high-mass binaries.

\begin{figure*}[ht]
    \centering
    \includegraphics[keepaspectratio,width=0.99\linewidth]{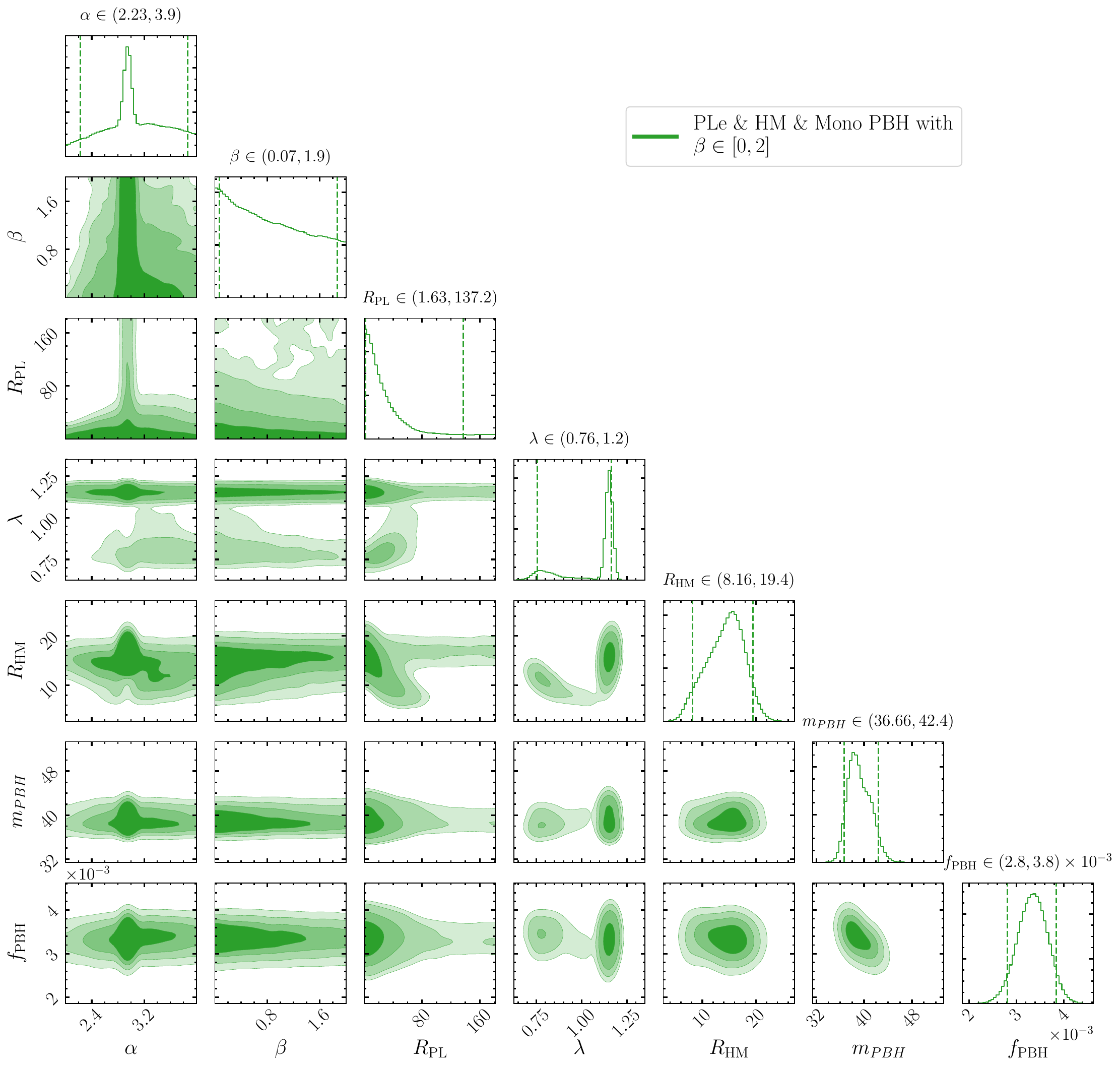}
    \vspace{-0.1in}
    \caption{Corner plot representing the fit parameter ranges for a population synthesis models that includes PLe $\&$ HM $\&$ monochromatic PBHs. $m_{\textrm{PBH}}$ is in $M_{\odot}$, while $f_{\textrm{PBH}}$ is unitless. In evaluating the range of values on $f_{\textrm{PBH}}$, we have assumed $f_{\textrm{PBH binaries}} = 0.5$.}
    \label{fig:PLe_HM_PBH_monochromatic_corner_plot}
\end{figure*}

In Fig.~\ref{fig:PLe_HM_PBH_monochromatic_corner_plot}, for the PLe $\&$ HM $\&$ Mono PBH population synthesis model, we show the seven-parameter fit ranges. 
Adding all three components reduces the best-fit merger rates of BBHs from the PLe component while it also allows for a higher best-fit merger rate of BBHs from HMs compared to the simpler PLe $\&$ HM model. 
The PBH component not only improves the fit but also allows for the HM population to better fit the $30-50$ $m_{1}$ mass range\footnote{We remind the reader that we account for mass resolution uncertainties in the LVK data as described in Section~\ref{subsec:SimulatingBinaries}, thus even a monochromatic PBH mass-distribution contributes in more than one mass-bin.}. 
While there is a preference for $\alpha \simeq 2.8-3.4$ and for $\beta \lesssim 1.0$, these two parameters are very weakly constrained. 
The $\lambda$ parameter in this model gets values in the $1.15^{+0.01}_{-0.39}$ range in closer agreement to Ref.~\cite{Ye:2025ano} than what we get for the PLe $\&$ HM model. 
A monochromatic PBH component with a mass in the range of 
$m_{\textrm{PBH}}=39.0^{+3.4}_{-2.3}$ is preferred for $f_{\textrm{PBH}} = (3.35^{+0.50}_{-0.53})\times 10^{-3}$ and can help explain the significant number of BBH detections with masses $\sim 40 M_{\odot}$. Those ranges are also given in Table~\ref{table:posteriors_pos_beta}.

Comparing the 
PLe $\&$ HM $\&$ Mono PBH population synthesis model to the PLe $\&$ HM  
we find a Bayes factor of $\textrm{BF}^{\textrm{PLe \& HM \& Mono PBH}}_{\textrm{PLe \& HM}} = 9$ and a likelihood ratio of $\Lambda^{\textrm{PLe \& HM \& Mono PBH}}_{\textrm{PLe \& HM}}= -36$; which suggest a significant improvement in the fit by adding a PBH population. 
We note that we also tested using a PL distribution instead of a PLe distribution for the first generation black holes. 
We get $\textrm{BF}^{\textrm{PL \& HM \& Mono PBH}}_{\textrm{PL \& HM}} = 13$ and a likelihood ratio of $\Lambda^{\textrm{PL \& HM \& Mono PBH}}_{\textrm{PL \& HM}}= -41$. 
The preference for a PBH component is present independently of the existence of an exponential cutoff on the $m_{1}$-distribution. 
Moreover, the preference for a third component like a PBH component is found as well using our wider priors given in Appendix~\ref{app:Alt_priors}.
The parameter fit ranges for the PL $\&$ HM $\&$ Mono PBH and the PL $\&$ HM population models are given in Table~\ref{table:posteriors_pos_beta}. 
In Figs.~\ref{fig:PLe_HM_corner_plot} and~\ref{fig:PLe_HM_PBH_monochromatic_corner_plot}, we chose to show the results for the PLe component, as even in these multi-component population models a PLe component for the first generation black holes is somewhat preferred to a PL one. 
We get  
$\textrm{BF}^{\textrm{PLe \& HM}}_{\textrm{PL \& HM}} = 4.9$ and $\textrm{BF}^{\textrm{PLe \& HM \& Mono PBH}}_{\textrm{PL \& HM \& Mono PBH}} = 1.5$. 

\begin{figure*}[ht]
    \centering
    \includegraphics[keepaspectratio,width=0.99\linewidth]{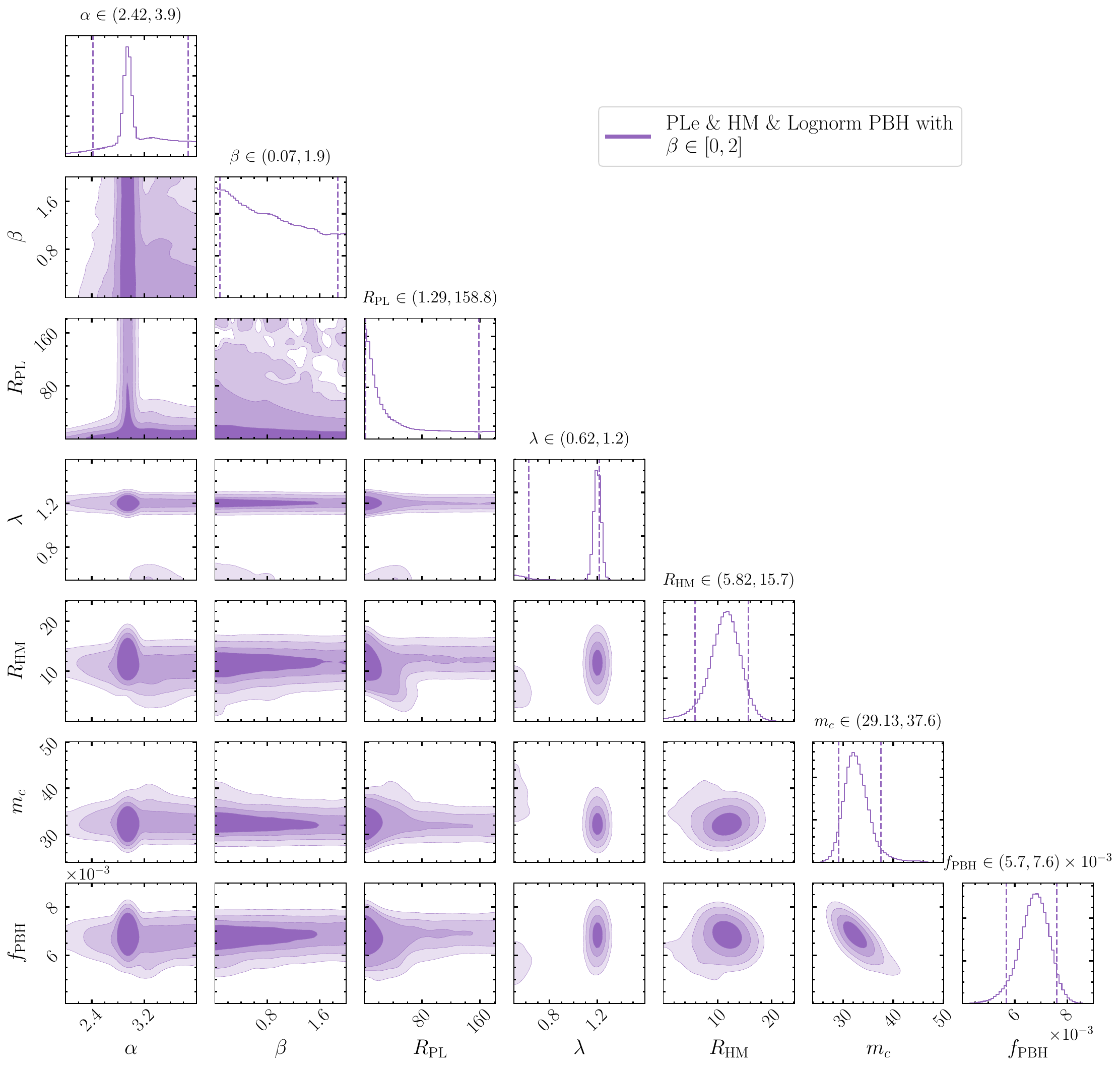}
    
    \vspace{-0.1in}
    \caption{Corner plot similar to Fig.~\ref{fig:PLe_HM_PBH_monochromatic_corner_plot}, but for the PLe $\&$ HM $\&$ Lognorm PBH population synthesis model. Again, in evaluating the range of values on $f_{\textrm{PBH}}$, we have assumed $f_{\textrm{PBH binaries}}= 0.5$.}
    \label{fig:PLe_HM_PBH_lognormal_corner_plot}
\end{figure*}

In Fig.~\ref{fig:PLe_HM_PBH_lognormal_corner_plot}, we replaced the assumption on the PBH mass-distribution. 
We show our results on the parameter ranges for the PLe $\&$ HM $\&$ Lognorm PBH population synthesis model. There is preference for $\alpha \simeq 2.7-3.1$ while $\beta \leq 1.5$, but as in the case of Fig.~\ref{fig:PLe_HM_PBH_monochromatic_corner_plot}, those two parameters are weakly constrained. Subsequently also the merger rate of the PLe component is weakly set to $R_{\textrm{PL}} = 23^{+136}_{-22}$. 
The HM component shows a more clear preference for $\lambda \simeq 1.2$ (with $\lambda = 1.20^{+0.02}_{-0.58}$ 90$\%$ CI) with a $R_{\textrm{HM}} = 11.4^{+4.2}_{-5.6}$. 
The lognormal PBH component is still well constrained to $m_{c} = 32.5^{+5.1}_{-3.4}$ and $f_{\textrm{PBH}} = 6.76^{+0.83}_{-1.07}$. 

Having a PBH component following a lognormal mass-distribution still has a clear statistical preference compared to not having a PBH component. 
We get $\textrm{BF}^{\textrm{PLe \& HM \& Lognorm PBH}}_{\textrm{PLe \& HM}} = 26$ and a likelihood ratio of $\Lambda^{\textrm{PLe \& HM \& Lognorm PBH}}_{\textrm{PLe \& HM}}= -66$ 
\footnote{When using instead the PL assumption for the first generation black holes we get $\textrm{BF}^{\textrm{PL \& HM \& Lognorm PBH}}_{\textrm{PL \& HM}} = 30$ and a likelihood ratio of $\Lambda^{\textrm{PL \& HM \& Lognorm PBH}}_{\textrm{PL \& HM}}= - 72$.}. 
In fact, the lognormal PBH distribution is clearly preferred to the monochromatic distribution by a 
$\textrm{BF}^{\textrm{PLe \& HM \& Lognorm PBH}}_{\textrm{PLe \& HM \& Mono PBH}} = 17$. 
Those two models have the same level of free parameters and prior ranges and moreover a lognormal mass-distribution is physically more motivated for PBHs (see Ref.~\cite{Aljaf:2024fru} and citations therein for a relevant discussion).
The lognormal PBH mass-distribution gives a wider range of masses around the observed peak of $40 \, M_{\odot}$ for the $m_{1}$ and $30 \, M_{\odot}$ for $m_{2}$, in agreement with the representation of the LVK data shown in Fig.~\ref{fig:m_1_2_z_hists} .

\section{Discussion and Conclusion} 
\label{sec:conc}

In this paper we have analyzed the LVK observations of BBH mergers from the O1 to the O4a observing runs (represented in Figs.~\ref{fig:m_1_2_z_hists} and~\ref{fig:m_1_z_distr}), in order to assess the origin of these merging binaries. We use \texttt{emcee} \cite{2013PASP..125..306F}, to test and maximize the log-likelihood of a given model with parameters within prior ranges, given the LVK observations (see Section~\ref{sebsec:Best_fit} and Table~\ref{table:priors}).

We test a sequence of population synthesis hypotheses. 
1) that these these binaries are just first generation black holes following a power-law mass distribution without or with an exponential cutoff in their primary mass (PL and PLe model respectively).
2) that in addition to that PL or PLe populations there is a component of hierarchical BH mergers (we refer to is as HM) possibly taking place in dense stellar environments. We use as a base model for that HM population the simulation work of Ref.~\cite{Ye:2025ano}, allowing for some degree of freedom in the predicted mass distribution of the second generation or higher BH masses (see Fig.~\ref{fig:hm_hists}). We refer to these population synthesis models as PL $\&$ HM or PLe $\&$ HM.
3) that in addition to the PL or PLe and the HM populations there is a third component of PBH merging binaries in the LVK observations. For that third component we test and derive limits for both the case that the PBH mass-distribution is monochromatic and the case that it arises from a lognormal distribution. For the PBH merger rates we rely on the latest estimates of Ref.~\cite{Aljaf:2025dta} (see also Fig.~\ref{fig:Total_rates:channels}).

Comparing the PL and PLe models to the LVK observations we find results similar to earlier works from the O1-O3 observations, i.e. a power-law index $\alpha \simeq 2-3$ with a mostly positive index $\beta$ for the distribution of the BBH $m_{2}/m_{1}$ ratio (see detailed discussion in Section~\ref{sec:method}). Depending on the exact prior assumptions on the range of $\alpha$ and $\beta$, the first generation BHs without any additional component of BBHs have a rate that is $R_{\textrm{PL}}\simeq 5-100$ $\textrm{Gpc}^{-3} \textrm{yr}^{-1}$ (Figs.~\ref{fig:PL_corner_plot} and~\ref{fig:PL_positive_beta_corner_plot}).

However, we find that the LVK observations strongly prefer a HM population in addition to the PL or PLe BBH population. We find that even more than half of the observed BBHs may be associated to the HM population (see Fig.~\ref{fig:M1_PLeandHM_histogram}), with a rate for that population to be $R_{\textrm{HM}}\simeq 5-20$ $\textrm{Gpc}^{-3} \textrm{yr}^{-1}$.
That can also reduce the rate of $R_{\textrm{PL}}$  to as low as  $ \sim  10 \, \textrm{Gpc}^{-3} \textrm{yr}^{-1}$ but in fact given the freedom in the other modeling parameters it makes the PL or PLe rates and the index $\beta$ mostly unconstrained as we show in Figs.~\ref{fig:PLe_HM_corner_plot},~\ref{fig:PLe_HM_negative_beta_corner_plot} and in Tables~\ref{table:posteriors_pos_beta} and~\ref{table:posteriors_neg_beta}. 
Furthermore, our fits indicate the need for a HM population that contains BHs that may even be third generation. To best-fit the observations we had to deform the mass-distribution of Ref.~\cite{Ye:2025ano}, in such a manner as to increase the masses of the black holes from HMs. 
Works as in Refs~\cite{Purohit:2024zkl, Li:2024zwt, Mai:2025jmk, 2026arXiv260204176N}, have also come to a similar conclusion. 
Whether a dominant HM population is indeed detected in gravitational waves, will be resolved by the continuing LVK and future observations. We leave such a question for future work. 

\begin{figure}
    %\centering
    \includegraphics[keepaspectratio,width=0.99\linewidth]{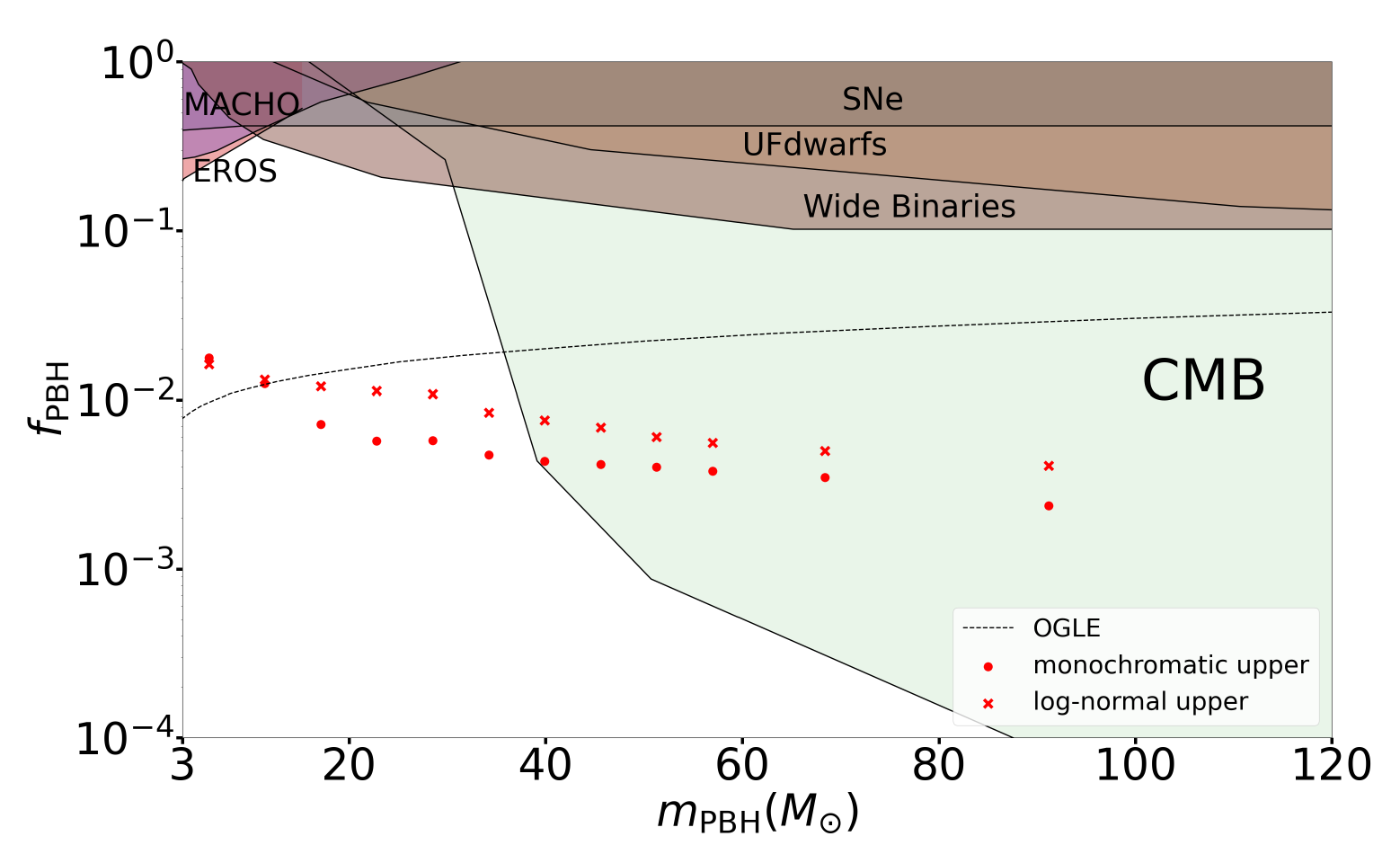}  
    \vspace{-0.3in}
    \caption{The $95\%$ upper limits on the PBH abundance $f_{\textrm{PBH}}$ vs the PBH mass $m_{\textrm{PBH}}$. In the dots we show the limits for a monochromatic mass-distribution, while with the ``x'' symbols the limits for a lognormal distribution with $m_{c}=m_{\textrm{PBH}}$. We assume $f_{\textrm{PBH binaries}} = 0.5$. We  provide for comparison upper limits from other probes on PBHs.}
    \label{fig:PBH_limits}
\end{figure}

Finally, we find that a small population of merging PBH binaries is favored to just the PL $\&$ HM or the PLe $\&$ HM synthesis models. Adding the PBH population improves the likelihood by $2 \Delta ln\mathcal{L}$ by $30 - 60$ and gives a Bayes factor of at least $ln \textrm{BF} = 15$ in favor of that third component to the observed BBHs (see Tables~\ref{table:posteriors_pos_beta} and~\ref{table:posteriors_neg_beta}, and the text in Section~\ref{sec:results}). 
This result has been reached with alternative prior assumptions. Whether from a monochromatic or a lognormal mass distribution, adding a PBH population helps better explain in the LVK observations the prominent peaks at $m_{1}\sim 40 \, M_{\odot}$ and $m_{2}\sim 30 \, M_{\odot}$ but also their redshift distribution. 
If such a population exists its preferred mass range is $\simeq 30-40 \, M_{\odot}$ with the 
PBHs accounting for a fraction $f_{\textrm{PBH}} \simeq 3-8 \times 10^{-3}$ of the observed 
dark matter abundance in the Universe (see Figs.~\ref{fig:PLe_HM_PBH_monochromatic_corner_plot},~\ref{fig:PLe_HM_PBH_lognormal_corner_plot} and~\ref{fig:PLe_HM_PBH_both_distributions_negative_beta_corner_plot}). In Fig.~\ref{fig:PBH_limits}, we present the updated limits on PBH abundance for masses between 5 and 80 $M_{\odot}$. The LVK limits are most competitive for masses in the range of $5-40 \, M_{\odot}$. We show other limits form Refs.~\cite{Zumalac_rregui_2018, Brandt_2016, Monroy_Rodr_guez_2014, Mr_z_2024, Mr_z_2024_2, Serpico_2020, Tisserand_2007, Alcock_2001, Carr_2021, Green:2020jor, PBH_bounds_zenodo}. The OGLE limits are depicted with a dashed line given their significant systematics (see  Ref.~\cite{Green:2025dut}). 

\begin{figure}[h!]
    \centering
\includegraphics[keepaspectratio,width=0.99\linewidth]{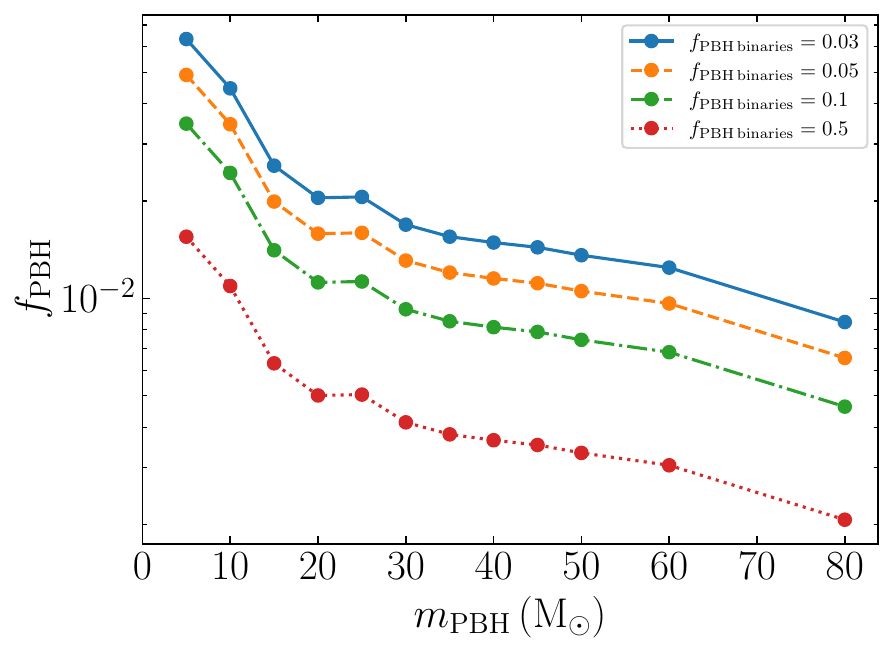}
    \caption{The $95\%$ upper limits on $f_{\rm PBH}$ as function of monochromatic PBH mass. These limits are obtained by fixing PBH binary fraction $f_{\textrm{PBH binaries}}$  at $0.03,0.05, 0.1$ and $0.5$. We observe that as we increase the value of $f_{\textrm{PBH binaries}}$, a significant suppression occurs in $f_{\rm PBH}$.}
    \label{fig:PBHlimits_with_fPBHbinaries}
\end{figure}

In the corner plots of Figs.~\ref{fig:PLe_HM_PBH_monochromatic_corner_plot},~\ref{fig:PLe_HM_PBH_lognormal_corner_plot} and~\ref{fig:PLe_HM_PBH_both_distributions_negative_beta_corner_plot}, and in Fig.~\ref{fig:PBH_limits} and in we assumed that half the PBHs are in binaries ($f_{\textrm{PBH binaries}} = 0.5$) which is closer to the typical assumption when comparing the gravitational-wave limits on PBHs to other probes.
However, for an $f_{\textrm{PBH}} \simeq 1\times 10^{-2}$ the fraction of PBHs in binaries is smaller than 0.5. 
In Fig.~\ref{fig:PBHlimits_with_fPBHbinaries}, we show the $95\%$ upper limits on $f_{\rm PBH}$ as a function of monochromatic PBH mass evaluated for several values on $f_{\textrm{PBH binaries}}$ of $0.03$, $0.05$, $0.1$, and $0.5$. Those values allow us to account for some level of uncertainty on the PBH binary population. Furthermore Ref.~\cite{Kavanagh:2018ggo}, has provided a scaling between $f_{\rm PBH}$ and $f_{\textrm{PBH binaries}}$ (see their Fig.~1). Accounting for that scaling, a more realistic limit on$f_{\rm PBH}$ with $m_{\textrm{PBH}}$ is provided for the lines with $f_{\textrm{PBH binaries}}$ 0.03-0.05, i.e. the range between the blue solid line and the orange dashed line. Thus for $m_{\textrm{PBH}}\simeq 30-40 \, M_{\odot}$, $f_{\textrm{PBH}} \sim 2\times 10^{-2}$. 

Future observations from ground-based gravitational-wave interferometers will allow us to further explore the properties of the population of first generation BHs, the contribution of hierarchical mergers as well as the potential contribution of PBHs in the observed BBH population. Including the spin of the observed black holes will also become an important tool in that endeavor (see for instance Ref~\cite{Berti:2025usa}). In particular since PBHs even in the most dense dark matter halos do not give hierarchical mergers of their own, if they come with a narrow mass range they would only contribute to the $m_{1}$ and $m_{2}$ mass spectra but have smaller typical values of spins compared to the HM population. Furthermore, the PBH comoving merger rates do increase with increasing redshift even to redshifts $z\sim 20$, when instead any HM population that requires time for this process to occur, could not have been prominent at too high redshifts. 

\begin{acknowledgments}
The authors would like to thank David Garfinkle, Iason Krommydas and Yi-Ming Zhong for valuable discussions.
MEB, MA and IC are supported by the National Science Foundation, under grant PHY-2207912. 

This research has made use of data or software obtained from the Gravitational Wave Open Science Center (gwosc.org), a service of the LIGO Scientific Collaboration, the Virgo Collaboration, and KAGRA. This material is based upon work supported by NSF's LIGO Laboratory which is a major facility fully funded by the National Science Foundation, as well as the Science and Technology Facilities Council (STFC) of the United Kingdom, the Max-Planck-Society (MPS), and the State of Niedersachsen/Germany for support of the construction of Advanced LIGO and construction and operation of the GEO600 detector. Additional support for Advanced LIGO was provided by the Australian Research Council. Virgo is funded, through the European Gravitational Observatory (EGO), by the French Centre National de Recherche Scientifique (CNRS), the Italian Istituto Nazionale di Fisica Nucleare (INFN) and the Dutch Nikhef, with contributions by institutions from Belgium, Germany, Greece, Hungary, Ireland, Japan, Monaco, Poland, Portugal, Spain. KAGRA is supported by Ministry of Education, Culture, Sports, Science and Technology (MEXT), Japan Society for the Promotion of Science (JSPS) in Japan; National Research Foundation (NRF) and Ministry of Science and ICT (MSIT) in Korea; Academia Sinica (AS) and National Science and Technology Council (NSTC) in Taiwan.
\end{acknowledgments}

\appendix

\appendix
\section{Testing alternative assumptions on the prior-ranges of models}
\label{app:Alt_priors}

In the main text we used flat priors in the ranges described by Table~\ref{table:priors}. Here we show results assuming $-1 \le \beta$.
Our main results are summarized in Table~\ref{table:posteriors_neg_beta} and can be directly compared to the results of Table~\ref{table:posteriors_pos_beta}.

\begin{table*}[t]
\centering
\begin{tabular}{| c | c | c | c | c | c | c | c | c | c | } 
 \hline
  Model & $\alpha$ & $\beta$ & $R_{\textrm{PL}}$ & $\lambda$ & $R_{\textrm{HM}}$ & $m_{\textrm{PBH}}$ & $f_{\textrm{PBH}}$ & $- 2 \Delta ln\mathcal{L}$ & $ln$BF \\ [0.8ex] 
   & &  & or $R_{\textrm{PLe}}$ & & & or $m_{c}$ & ($\times 10^{-3}$)  &  & \\ 
 \hline
 PL & $2.80^{+0.15}_{-0.11}$ & $0.83^{+1.45}_{-1.11}$ & $121^{+78}_{-29}$ & - & - & - & - & 0 & 0\\
 \hline
 PLe & $2.04^{+0.09}_{-0.04}$ & $1.19^{+0.74}_{-0.90}$ & $73^{+21}_{-11}$ & - & - & - & - & -40 & +21 \\
 \hline
 PL $\&$ HM & $3.43^{+0.46}_{-0.69}$ & $-0.08^{+2.05}_{-0.85}$ & $77^{+104}_{-74}$ & $1.45^{+0.05}_{-0.42}$ & $7.7^{+11.5}_{-1.6}$ & - & - & -141 & +66\\
 \hline
 PLe $\&$ HM & $2.49^{+0.39}_{-0.33}$ & $-0.06^{+1.66}_{-0.63}$ & $74^{+92}_{-37}$ & $1.48^{+0.02}_{-0.07}$ & $5.9^{+1.5}_{-1.2}$ & - & - & -151 & +71\\
 \hline
PL $\&$ HM $\&$ Mono PBH  &  $3.54^{+0.41}_{-0.84}$ & $-0.30^{+2.33}_{-0.66}$ & $39^{+126}_{-37}$ & $0.85^{+0.31}_{-0.14}$ & $12.2^{+5.9}_{-4.4}$ & $38.5^{+3.3}_{-2.2}$ & $3.41^{+0.48}_{-0.53}$  & -185 & +79\\
 \hline
 PLe $\&$ HM $\&$ Mono PBH  & $2.99^{+0.89}_{-0.82}$ & $-0.25^{+1.94}_{-0.70}$ & $40^{+118}_{-36}$ & $1.15^{+0.01}_{-0.40}$ & $14.4^{+4.8}_{-6.3}$ & $38.9^{+3.4}_{-2.2}$ & $3.35^{+0.51}_{-0.53}$ & -186 & +81 \\
 \hline
 PL $\&$ HM $\&$ Lognorm PBH  &  $3.52^{+0.43}_{-0.99}$ & $-0.42^{+2.48}_{-0.56}$ & $24^{+137}_{-23}$ & $1.19^{+0.02}_{-0.67}$ & $10.7^{+4.3}_{-6.3}$ & $32.8^{+6.0}_{-3.6}$ & $6.67^{+0.87}_{-1.24}$ & -216 & +96 \\
 \hline
 PLe $\&$ HM $\&$ Lognorm PBH  & $3.04^{+0.86}_{-0.69}$ & $-0.30^{+2.00}_{-0.66}$ & $42^{+121}_{-40}$ & $1.20^{+0.02}_{-0.63}$ & $11.2^{+4.4}_{-5.6}$ & $32.6^{+4.6}_{-3.4}$ & $6.72^{+0.83}_{-0.99}$ & -216 & +97 \\
 \hline
\end{tabular}
\caption{The fit ranges for the tested models. We present here the results for the prior assumption that $\beta \in [-1, 2]$ (we give the results for the alternative prior on $\beta$ in Appendix~\ref{app:Alt_priors}).
The central values give the median value from the posterior PDF. The upper and lower give the $90\%$ credible interval range that are also represented in the relevant corner plots as dashed lines.
Parameters $\alpha$, $\beta$, $\lambda$ and $f_{\textrm{PBH}}$ are without units, $R_{\textrm{PL}}$ and $R_{\textrm{HM}}$ are in $\textrm{Gpc}^{-3} \textrm{yr}^{-1}$ and $m_{\textrm{PBH}}$ or $m_{c}$ are in $M_{\odot}$.
The second to last column gives the $-2\Delta ln \mathcal{L}$ between any population synthesis model ``$X$'' and the simple PL model, i.e. $\Lambda^{X}_{PL}$. The last column gives in turn the logarithm of the Bayes factor $ln\textrm{BF}^{X}_{\textrm{PL}}$ in comparing model ``$X$'' to the PL model.}
\label{table:posteriors_neg_beta}
\end{table*}

\begin{figure*}[ht]
    \centering
    \includegraphics[keepaspectratio,width=0.85\linewidth]
    {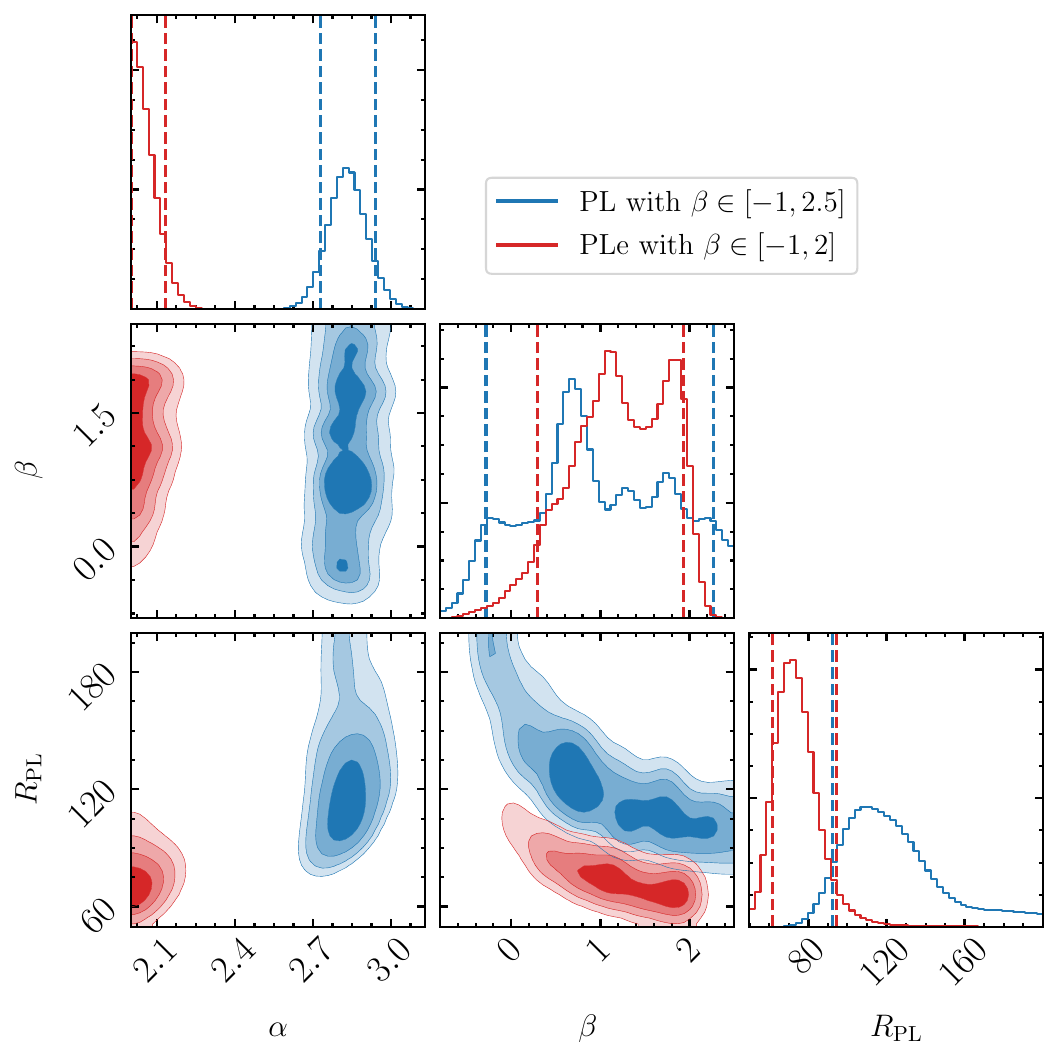}
    \vspace{-0.1in}
    \caption{Corner plot similar to that of Fig.~\ref{fig:PL_corner_plot}, but with prior assumption on $-1 \leq \beta$.}
    \label{fig:PL_positive_beta_corner_plot}
\end{figure*}

We find that for the simple PL and PLe population synthesis models the posterior PDFs and the $90 \%$ CI ranges of $\alpha$ are essentially the same (the upper and lower values defining them change within a few $\%$ between the $-1 \leq \beta$ and the $0 \leq \beta$ priors)\footnote{We remind the reader that for the PL model we used $\beta \leq 2.5$ and for the PLe $\beta \leq 2$.}. Also, the results for  $R_{\textrm{PL}}$ are similar, though spanning a somewhat less constrained range; while the result for the upper end of the $\beta$ range are roughly the same as well. In Fig.~\ref{fig:PL_positive_beta_corner_plot}, show for the PL and the PLe populations our equivalent results to those presented in Fig.~\ref{fig:PL_corner_plot} of the main text.

\begin{figure*}[ht]
    \centering
    \includegraphics[keepaspectratio,width=0.9\linewidth]{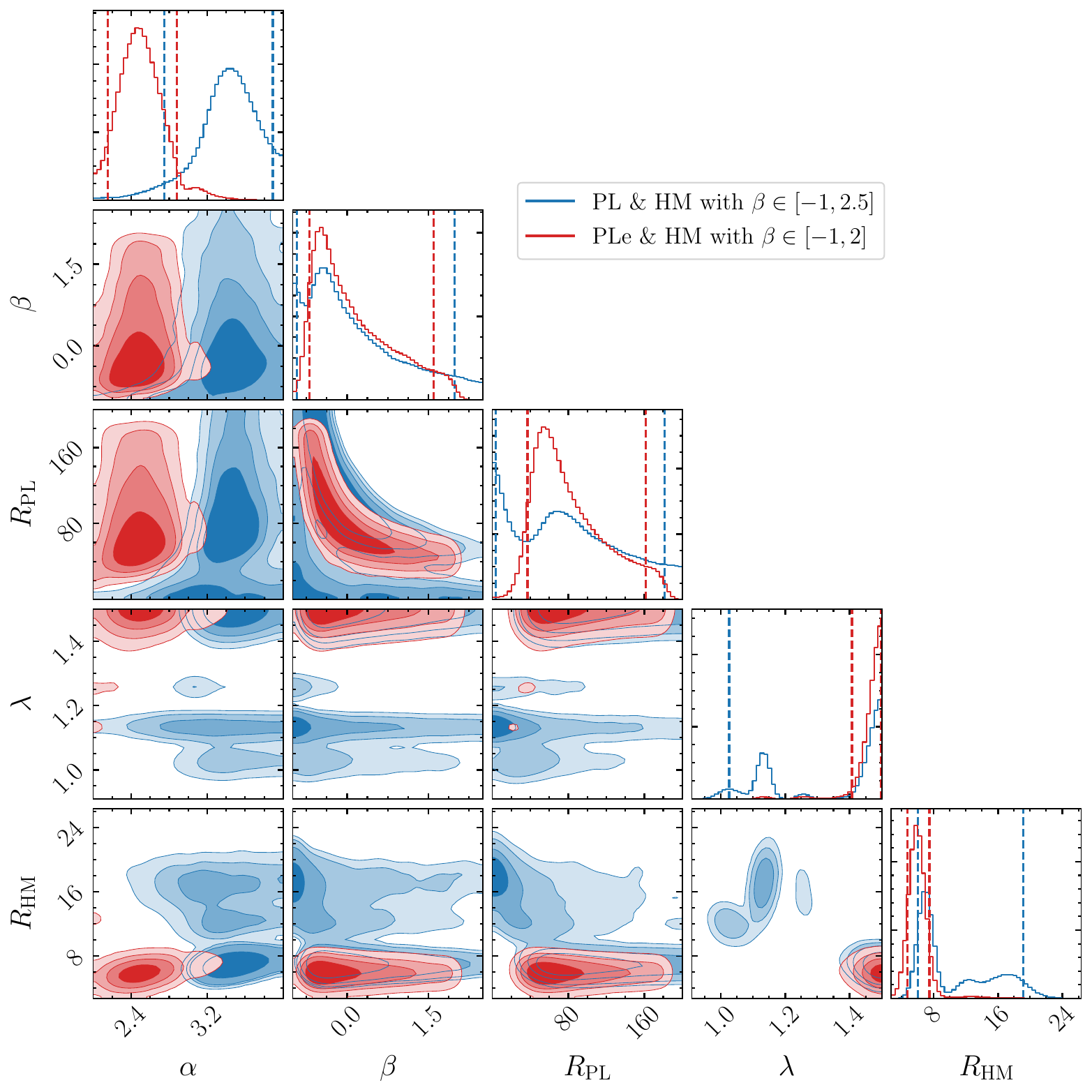}
    \vspace{-0.1in}
    \caption{Corner plot similar to that of Fig.~\ref{fig:PLe_HM_corner_plot}, but with prior assumption on $- 1 \leq \beta$.}
    \label{fig:PLe_HM_negative_beta_corner_plot}
\end{figure*}

For the PL $\&$ HM and the PLe $\&$ HM our fit results are shown in Fig.~\ref{fig:PLe_HM_negative_beta_corner_plot}. 
The alternative choice for the $\beta$ priors has a marginal effect on the contour ranges and the $90 \%$ CIs  (changes of $O(10\%)$ of the values that define the range of the CIs) for $\alpha$, $\lambda$ and the $R_{\textrm{HM}}$. 
Changing the prior range for the $\beta$ index only affects its posterior range and importantly the $R_{\textrm{PL}}$ range. Allowing the $\beta$ parameter to be negative, lets the merger rates of the PL and PLe components to take larger values. 
This is expected. As we described in the main text, negative values of $\beta$ allow for many BBHs in the Universe with low values of mass ratios, many of which would go undetected.

\begin{figure*}[t]
    \centering
   \includegraphics[keepaspectratio,width=0.60\linewidth]{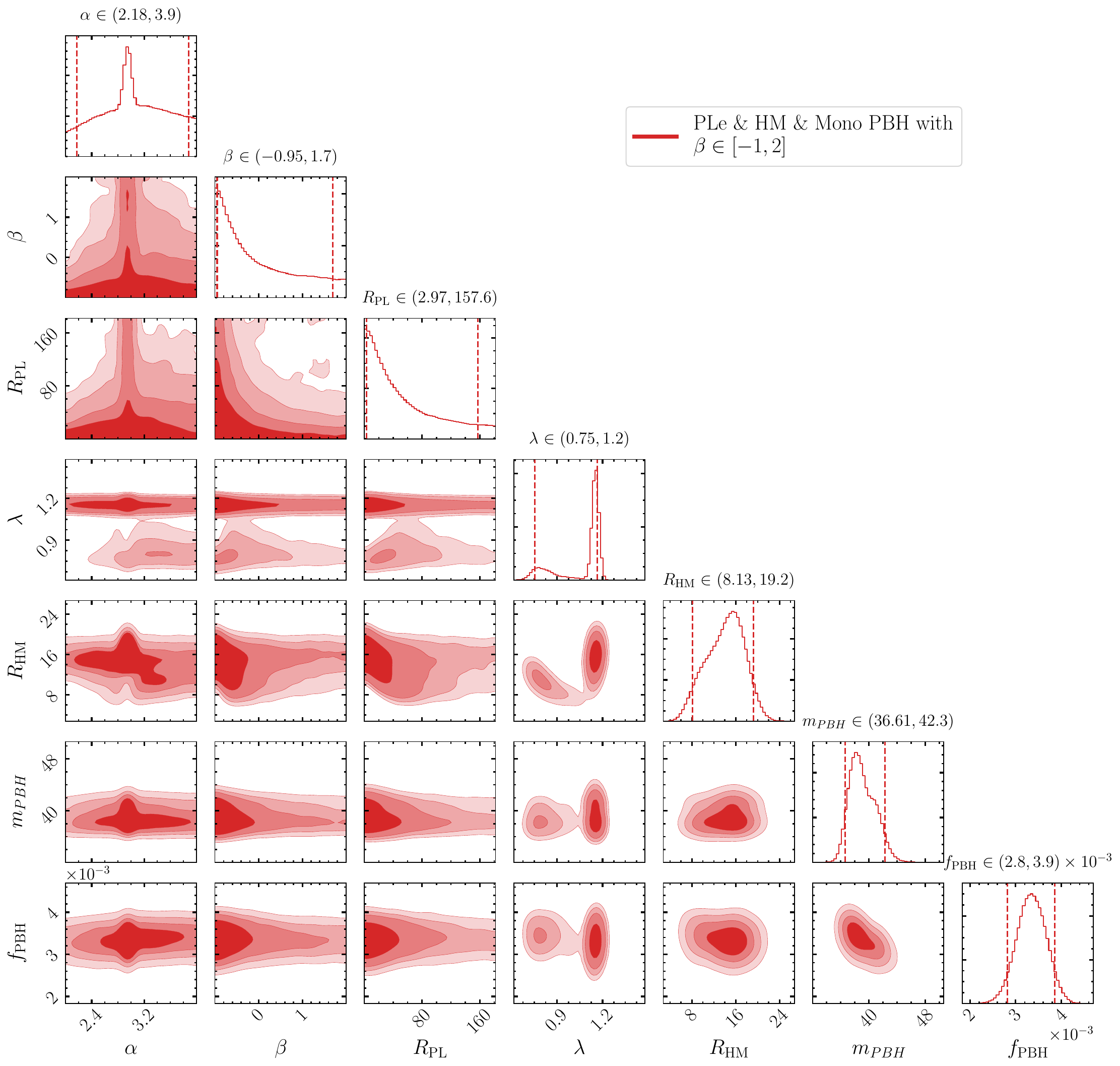}\\
    \includegraphics[keepaspectratio,width=0.60\linewidth]{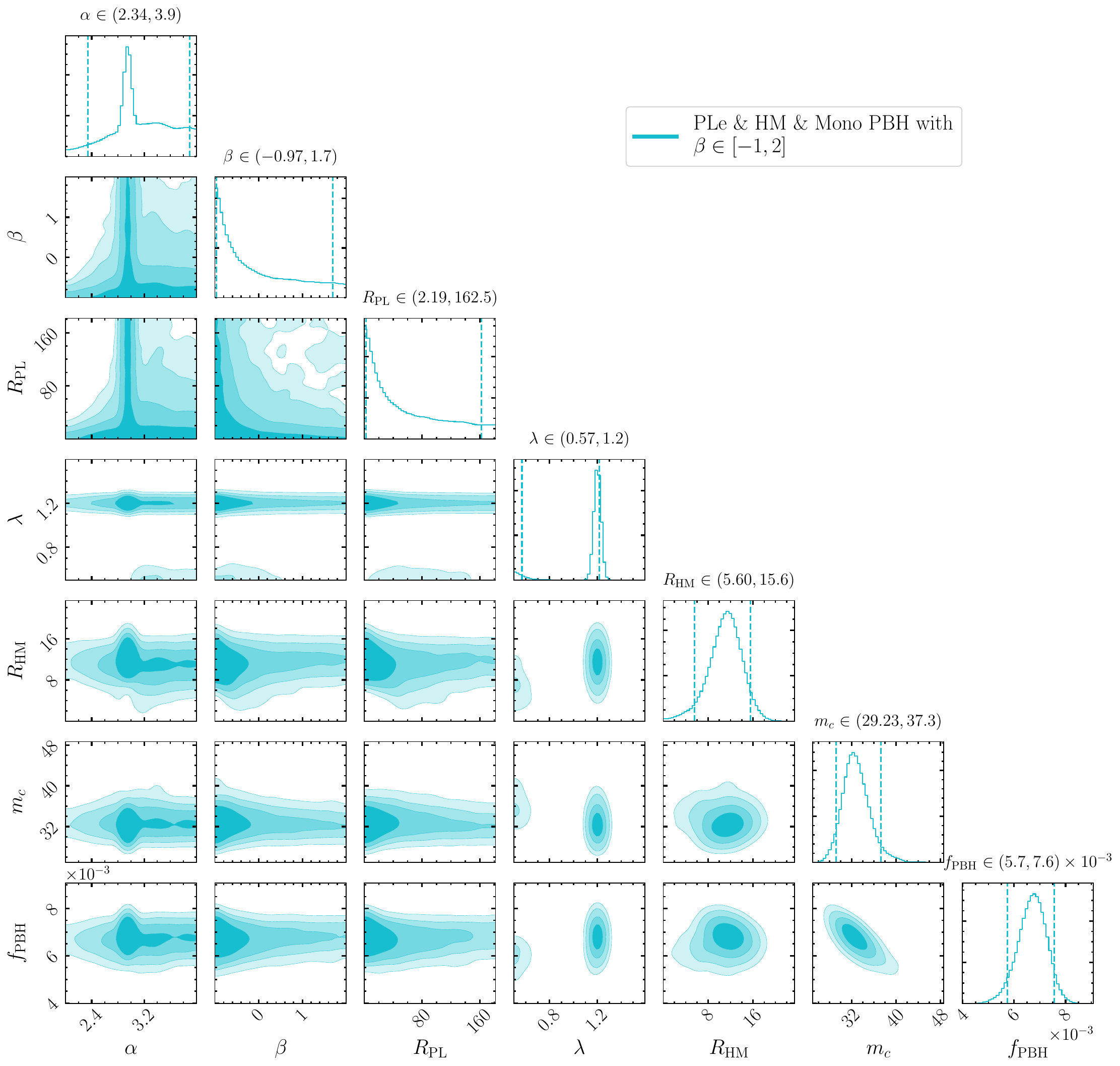}
    \vspace{-0.1in}
    \caption{Corner plot as in Fig.~\ref{fig:PLe_HM_PBH_monochromatic_corner_plot}, (top panel) and Fig.~\ref{fig:PLe_HM_PBH_lognormal_corner_plot}, (bottom panel)  representing the fit parameter values for PLe $\&$ HM $\&$ Mono PBH (top) and PLe $\&$ HM $\&$ Lognorm PBH (bottom) population synthesis models, for a prior on $-1 \leq \beta$.}
    \label{fig:PLe_HM_PBH_both_distributions_negative_beta_corner_plot}
\end{figure*}

In Fig.~\ref{fig:PLe_HM_PBH_both_distributions_negative_beta_corner_plot}, we show our fit results for the PLe $\&$ HM $\&$ Mono PBH population model (top panel) and for the PLe $\&$ HM $\&$ Lognorm PBH model (bottom panel). 
Our results for the PBH properties, i.e their mass-distribution and their abundance $f_{\textrm{PBH}}$ are essentially the same to those from the main text. 
Also, the ranges for the index $\alpha$ the the HM population's $\lambda$ parameter and their merger rates $R_{\textrm{HM}}$ are very similar to the results given in Table~\ref{table:posteriors_pos_beta} and Figs.~\ref{fig:PLe_HM_PBH_monochromatic_corner_plot} and~\ref{fig:PLe_HM_PBH_lognormal_corner_plot}.
Again the main effect of increasing the range of the prior distribution of $\beta$ is to expand its posterior distribution to lower values as well and to increase somewhat the $90 \%$ CI ranges for rate $R_{\textrm{PL}}$. 

Finally, for our tests with the $-1 \leq \beta$ priors, we find very similar differences in the log-likelihoods $- 2 \Delta ln\mathcal{L}$ and also in the evidence values for these populations and subsequently the Bayes factors when comparing between alternative population synthesis models. 
The statements made in the main text regarding the preference for a HM population to a simple PL or a PLe model and further statistical preference for an additional PBH component to the PL $\&$ HM or the PLe $\&$ HM models remain true.  In fact, in Table~\ref{table:posteriors_pos_beta_LargeRPL}, we show results where we let $R_{\textrm{PL}}$ and $R_{\textrm{PLe}}$ to take as high values as $2\times 10^{3} \, \textrm{Gpc}^{-3} \textrm{yr}^{-1}$. Even for that extremely free set of priors for the PL and PLe rates there still is preference for a HM and a PBH population with effectively the same values on their model parameters.   

\begin{table*}[ht]
\centering
\begin{tabular}{| c | c | c | c | c | c | c | c | c | } 
 \hline
  Model & $\alpha$ & $\beta$ & $R_{\textrm{PL}}$ & $\lambda$ & $R_{\textrm{HM}}$ & $m_{\textrm{PBH}}$ & $f_{\textrm{PBH}}$ & $- 2 \Delta ln\mathcal{L}$ \\ [0.8ex] 
   & &  & or $R_{\textrm{PLe}}$ & & & or $m_{c}$ &  ($\times 10^{-3}$) & \\ 
 \hline
 PL & $2.84^{+0.11}_{-0.10}$ & $1.11^{+1.17}_{-0.89}$ & $117^{+37}_{-25}$ & - & - & - & - & 0\\
 \hline
 PLe & $2.04^{+0.09}_{-0.04}$ & $1.19^{+0.74}_{-0.83}$ & $73^{+19}_{-11}$ & - & - & - & - & -40 \\
 \hline
 PL $\&$ HM & $3.46^{+0.43}_{-0.51}$ & $0.79^{+1.44}_{-0.73}$ & $63^{+50}_{-59}$ & $1.46^{+0.03}_{-0.44}$ & $7.3^{+10.8}_{-1.3}$ & - & - & -140 \\
 \hline
 PLe $\&$ HM & $2.51^{+0.38}_{-0.34}$ & $0.66^{+1.15}_{-0.61}$ & $54^{+29}_{-24}$ & $1.48^{+0.02}_{-0.19}$ & $5.9^{+2.4}_{-1.2}$ & - & - & -150 \\
 \hline
 PL $\&$ HM $\&$ Mono PBH  & $3.53^{+0.43}_{-0.86}$ & $0.95^{+1.36}_{-0.87}$ & $15^{+48}_{-14}$ & $1.14^{+0.02}_{-0.42}$ & $13.4^{+5.4}_{-5.3}$ & $38.7^{+3.5}_{-2.2}$ & $ 3.38^{+0.50}_{-0.52}$ & -185 \\
 \hline
 PLe $\&$ HM $\&$ Mono PBH  &   $2.96^{+0.75}_{-0.40}$ & $0.93^{+0.96}_{-0.85}$ & $412^{+1261}_{-406}$ & $1.15^{+0.01}_{-0.41}$ & $15.7^{+4.4}_{-7.3}$ & $38.8^{+3.5}_{-2.2}$ & $3.37^{+0.50}_{-0.55}$& -186 \\
 
 \hline
 PL $\&$ HM $\&$ Lognorm PBH  &  $3.53^{+0.42}_{-0.92}$ & $0.99^{+1.34}_{-0.91}$ & $8.4^{+47.7}_{-8.0}$ & $1.19^{+0.02}_{-0.67}$ & $10.8^{+4.4}_{-6.3}$ & $32.7^{+5.2}_{-3.5}$ & $6.70^{+0.85}_{-1.16}$& -216\\
 \hline
 PLe $\&$ HM $\&$ Lognorm PBH  &  $2.96^{+0.65}_{-0.06}$ & $0.98^{+0.92}_{-0.88}$ & $692^{+1013}_{-686}$ & $1.20^{+0.02}_{-0.63}$ & $12.0^{+4.2}_{-5.7}$ & $32.3^{+5.0}_{-3.3}$ & $ 6.83^{+0.81}_{-1.06}$ & -216 \\
 \hline
\end{tabular}
\caption{The fit ranges for the tested models. We present here the results for the prior assumption that $0 \leq \beta$ with $R_{\textrm{PL}}$ and $R_{\textrm{PLe}} \leq 2\times 10^{3} \textrm{Gpc}^{-3} \textrm{yr}^{-1}$.
As in Table~\ref{table:posteriors_pos_beta}, the central values give the median value from the posterior PDF of each parameter. The upper and lower give the $90\%$ credible interval range that are also represented in the relevant corner plots as dashed lines. The last column gives the $-2\Delta ln \mathcal{L}$ between any population synthesis model ``$X$'' and the simple PL model. While the rates $R_{\textrm{PL}}$ and $R_{\textrm{PLe}}$ span much greater ranges, compared to the results of the main text, our results for the HM and the PBH components properties and statistical significance are effectively unchanged. }
\label{table:posteriors_pos_beta_LargeRPL}
\end{table*}
%\clearpage
\bibliography{LIGO_O4a.bib}%{}
% \bibliography{main.bib}{}
% \bibliographystyle{apsrev4-2}
\end{document}